\documentclass[preprint,final,11pt]{elsarticle}
\usepackage{lineno}
\newcommand*\patchAmsMathEnvironmentForLineno[1]{%
  \expandafter\let\csname old#1\expandafter\endcsname\csname #1\endcsname
  \expandafter\let\csname oldend#1\expandafter\endcsname\csname end#1\endcsname
  \renewenvironment{#1}%
     {\linenomath\csname old#1\endcsname}%
     {\csname oldend#1\endcsname\endlinenomath}}%
\newcommand*\patchBothAmsMathEnvironmentsForLineno[1]{%
  \patchAmsMathEnvironmentForLineno{#1}%
  \patchAmsMathEnvironmentForLineno{#1*}}%
\AtBeginDocument{%
\patchBothAmsMathEnvironmentsForLineno{equation}%
\patchBothAmsMathEnvironmentsForLineno{align}%
\patchBothAmsMathEnvironmentsForLineno{flalign}%
\patchBothAmsMathEnvironmentsForLineno{alignat}%
\patchBothAmsMathEnvironmentsForLineno{gather}%
\patchBothAmsMathEnvironmentsForLineno{multline}%
}

\usepackage{microtype}
\usepackage{hyperref}

\usepackage{amssymb}

\usepackage{array}
\usepackage{fixltx2e}
\usepackage{graphicx}
\usepackage{epstopdf}
\graphicspath{{figs_pdf/}}
\usepackage{amsmath}
\usepackage{setspace}
\usepackage{enumerate}
\usepackage[ruled,linesnumbered,vlined]{algorithm2e}
\usepackage{color}
\usepackage{array}
\newcolumntype{C}[1]{>{\centering\let\newline\\\arraybackslash\hspace{0pt}}m{#1}}

\journal{}

\begin{document}

\begin{frontmatter}



\title{Efficient Feature-based Image Registration by Mapping Sparsified Surfaces}

\author[label1]{Chun Pang Yung}
\author[label2]{Gary P. T. Choi}
\author[label3]{Ke Chen}
\author[label1]{Lok Ming Lui\corref{cor1}}

\address[label1]{Department of Mathematics, The Chinese University of Hong Kong, Hong Kong}
\address[label2]{John A. Paulson School of Engineering and Applied Sciences, Harvard University, USA}
\address[label3]{Department of Mathematical Sciences, The University of Liverpool, United Kingdom}
\cortext[cor1]{Corresponding author.}
\ead{lmlui@math.cuhk.edu.hk}

\begin{abstract}
With the advancement in the digital camera technology, the use of high resolution images and videos has been widespread in the modern society. In particular, image and video frame registration is frequently applied in computer graphics and film production. However, conventional registration approaches usually require long computational time for high resolution images and video frames. This hinders the application of the registration approaches in the modern industries. In this work, we first propose a new image representation method to accelerate the registration process by triangulating the images effectively. For each high resolution image or video frame, we compute an optimal coarse triangulation which captures the important features of the image. Then, we apply a surface registration algorithm to obtain a registration map which is used to compute the registration of the high resolution image. Experimental results suggest that our overall algorithm is efficient and capable to achieve a high compression rate while the accuracy of the registration is well retained when compared with the conventional grid-based approach. Also, the computational time of the registration is significantly reduced using our triangulation-based approach.
\end{abstract}

\end{frontmatter}


\section{Introduction}

In recent decades, the rapid development of the digital camera hardware has revolutionized human lives. On one hand, even mid-level mobile devices can easily produce high resolution images and videos. Besides the physical elements, the widespread use of the images and videos also reflects the importance of developing software technology for them. On the other hand, numerous registration techniques for images and video frames have been developed for a long time. The existing registration techniques work well on problems with a moderate size. However, when it comes to the current high quality images and videos, most of the current registration techniques suffer from extremely long computations. This limitation in software seriously impedes fully utilizing the state-of-the-art camera hardware.

One possible way to accelerate the computation of the registration is to introduce a much coarser grid on the images or video frames. Then, the registration can be done on the coarse grid instead of the high resolution images or video frames. Finally, the fine details can be added back to the coarse registration. It is noteworthy that the quality of the coarse grid strongly affects the quality of the final registration result. If the coarse grid cannot capture the important features of the images or video frames, the final registration result is likely to be unsatisfactory. In particular, for the conventional rectangular coarse grids, since the partitions are restricted in the vertical and horizontal directions, important features such as slant edges and irregular shapes cannot be effectively recorded. By contrast, triangulations allow more freedom in the partition directions as well as the partition sizes. Therefore, it is more desirable to make use of triangulations in simplifying the registration problems.

In this work, we propose a two-stage algorithm for effective registration of specially large images. In stage 1, a content-aware image representation algorithm to {\em TR}iangulate {\em IM}ages, abbreviated as \emph{TRIM}, is developed to simplify high quality images and video frames. Specifically, for each high quality image or video frame, we compute a coarse triangulation representation of it. The aim is to create a high quality triangulation  on the set of the content-aware sample points using the Delaunay triangulation. The computation involves a series of steps including subsampling, unsharp masking, segmentation and sparse feature extraction for locating sample points on important features. Then in stage 2, using coarse triangular representation of the images, the registration is computed by a landmark-based quasi-conformal registration algorithm \cite{Lam14} for computing the coarse registration. The fine detail of the image or video frame in high resolution is computed with the aid of a mapping interpolation. Our proposed framework may be either used as a standalone fast registration algorithm or also served as a highly efficient and accurate initialization for other registration approaches.

The rest of this paper is organized as follows. In Section \ref{previous}, we review the literature on image and triangular mesh registration. Our proposed method is explained in details in Section \ref{main}. In Section \ref{experiment}, we demonstrate the effectiveness of our approach with numerous real images. The paper is concluded in Section \ref{conclusion}.

\section{Previous works} \label{previous}

In this section, we describe the previous works closely related to our work.

Image registration have been widely studied by different research groups. Surveys on the existing image registration approaches can be found in \cite{Zitova03,Crum04,Klein09,ZJ17}. In particular, one common approach for guaranteeing the accuracy of the registration is to make use of landmark constraints. Bookstein \cite{Bookstein78,Bookstein91,Bookstein99} proposed the unidirectional landmark thin-plate spline (UL-TPS) image registration. In \cite{Johnson02}, Johnson and Christensen presented a landmark-based consistent thin-plate spline (CL-TPS) image registration algorithm. In \cite{Joshi00}, Joshi et al. proposed the Large Deformation Diffeomorphic Metric Mapping (LDDMM) for registering images with a large deformation. In \cite{Glaunes04,Glaunes08}, Glaun\`es et al. computed large deformation diffeomorphisms of images with prescribed displacements of landmarks.

A few works on image triangulations have been reported. In \cite{Gee94}, Gee et al. introduced a probabilistic approach to the brain image matching problem and described the finite element implementation. In \cite{Kaufmann13}, Kaufmann et al. introduced a framework for image warping using the finite element method. The triangulations are created using the Delaunay triangulation method \cite{Shewchuk96} on a point set distributed according to variance in saliency. In \cite{Lehner07,Lehner08}, Lehner et al. proposed a data-dependent triangulation scheme for image and video compression. Recently, Yun \cite{dmesh} designed a triangulation image generator called DMesh based on the Delaunay triangulation method \cite{Shewchuk96}.

In our work, we handle image registration problems with the aid of triangulations. Numerous algorithms have been proposed for the registration of triangular meshes. In particular, the landmark-driven approaches use prescribed landmark constraints to ensure the accuracy of mesh registration. In \cite{Wang05,Lui07,Wang07}, Wang et al. proposed a combined energy for computing a landmark constrained optimized conformal mapping of triangular meshes. In \cite{Lui10}, Lui et al. used vector fields to represent surface maps and computed landmark-based close-to-conformal mappings. Shi et al. \cite{Shi13} proposed a hyperbolic harmonic registration algorithm with curvature-based landmark matching on triangular meshes of brains. In recent years, quasi-conformal mappings have been widely used for feature-endowed registration \cite{Zeng11,Zeng14,Lui14,Meng16}. Choi et al. \cite{Choi15} proposed the FLASH algorithm for landmark aligned harmonic mappings by improving the algorithm in \cite{Wang05,Lui07} with the aid of quasi-conformal theories. In \cite{Lam14}, Lam and Lui reported the quasi-conformal landmark registration (QCLR) algorithm for triangular meshes.

{\bf Contributions}.
Our proposed  approach for fast registration of high resolution images or video frames is advantageous in the following aspects:
\begin{enumerate}[(1).]
 \item The triangulation algorithm is fully automatic. The important features of the input image are well recorded in the resulting coarse triangulation.

 \item The algorithm is fast and robust. The coarse triangulation of a typical high resolution image can be computed within seconds.

 \item The registration algorithm for the triangulated surfaces by a Beltrami framework incorporates both the edge and landmark constraints to deliver a better quality map as fine details are restored. By contrast, for regular grid-based approaches, the same landmark correspondences can only be achieved on the high resolution image representation.

 \item Using our approach, the problem scale of the image and video frame registration is significantly reduced. Our method can alternatively serve as a fast and accurate initialization for the state-of-the-art image registration algorithms.
 
\end{enumerate}

\begin{figure}[t!]
 \centering
 \includegraphics[width=0.65\textwidth]{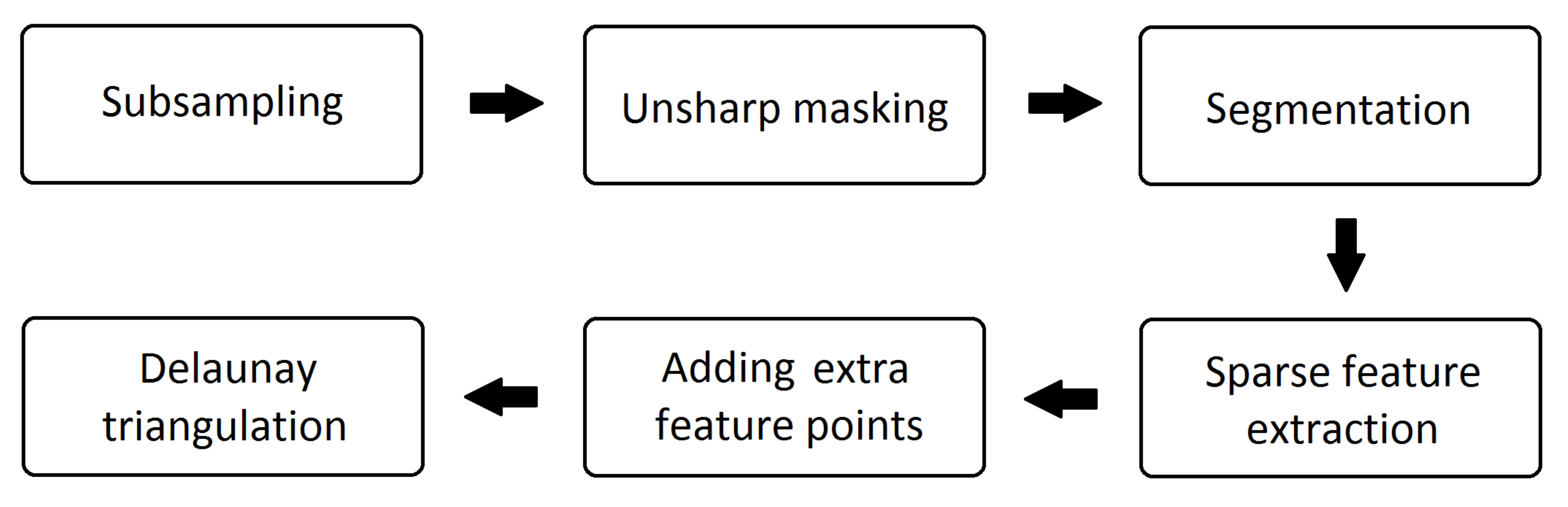}
 \caption{The pipeline of our proposed {\em TRIM} algorithm for accelerating image registration via coarse triangulation.}
 \label{fig:pl}
\end{figure}

\section{Proposed method} \label{main}

In this section, we describe our proposed approach for efficient image registration  in details.

\subsection{Stage $1$ -- Construction of coarse triangulation on images}
Given two high resolution images $I_1$ and $I_2$, our goal is to compute a fast and accurate mapping $f: I_1 \to I_2$. Note that directly working on the high resolution images can be inefficient. To accelerate the computation, the first step is to construct a coarse triangular representation of the image $I_1$. In the following, we propose an efficient image triangulation scheme called {\em TRIM}. The pipeline of our proposed framework is described in Figure \ref{fig:pl}.

Our triangulation scheme is content-aware. Specifically, special objects and edges in the images are effectively captured by a segmentation step, and a suitable coarse triangulation is constructed with the preservation of these features. Our proposed {\em TRIM} method consists of 6 steps in total.

\subsubsection{Subsampling the input image without affecting the triangulation quality}
Denote the input image by $I$. To save the computational time for triangulating the input image $I$, one simple remedy is to reduce the problem size by performing certain subsampling on $I$. For ordinary images, subsampling unavoidably creates adverse effects on the image quality. Nevertheless, it does not affect the quality of the coarse triangulation we aim to construct on images. 

In our triangulation scheme, we construct triangulations based on the straight edges and special features on the images. Note that straight edges are preserved in all subsamplings of the images because of the linearity. More specifically, if we do subsampling on a straight line, the subsampled points remain to be collinear. Hence, our edge-based triangulation is not affected by the mentioned adverse effects. In other words, we can subsample high resolution images to a suitable size for enhancing the efficiency of the remaining steps for the construction of the triangulations. We denote the subsampled image by $\tilde{I}$. In practice, for images larger than $1000\times 1000$, we subsample the image so that it is smaller than $1000\times 1000$.

\subsubsection{Performing unsharp masking on the subsampled image}
After obtaining the subsampled image $\tilde{I}$, we perform an unsharp masking on $\tilde{I}$ in order to preserve the edge information in the final triangulation. More specifically, we first transform the data format of the subsampled image $\tilde{I}$ to the CIELAB standard. Then, we apply the unsharp masking method in \cite{Polesel00} on the intensity channel of the CIELAB representation of $\tilde{I}$. The unsharp masking procedure is briefly described as follows.

By an abuse of notation, we denote $\tilde{I}(x,y)$ and $\bar{I}(x,y)$ the intensities of the input subsampled image $\tilde{I}$ and the output image $\bar{I}$ respectively, and $G_{\sigma}(x,y)$ the Gaussian mean of the intensity of the pixel $(x,y)$ with standard derivation $\sigma$. Specifically, $G_{\sigma}(x,y)$ is given by
\begin{equation}
 G_{\sigma}(x,y) \triangleq \frac{1}{\sigma\sqrt{2\pi}}\int_{(u,v) \in \Omega} e^{-\frac{(u-x)^2 + (v-y)^2}{2\sigma^2}}.
\end{equation}
We perform an unsharp masking on the image using the following formula
\begin{equation}
 \bar{I}(x,y) = \tilde{I}(x,y) - \lambda\begin{cases} G_{\sigma}*\tilde{I}(x,y) & \text{ if } V_s(x,y)>\theta, \\ 0 & \text{ if } V_s(x,y) < \theta, \end{cases}
\end{equation}
where
\begin{equation}
 V_s(x,y) \triangleq \sqrt{\frac{1}{Area(M_s)} \int_{(u,v) \in M_s} (\tilde{I}(u,v)-\tilde{I}_{mean}(x,y))^2 }
\end{equation}
and
\begin{equation}
\tilde{I}_{mean}(x,y) = \frac{1}{Area(M_s)} \int_{(u,v) \in M_s} \tilde{I}(u,v).
\end{equation}

\begin{figure}[t]
 \centering
 \includegraphics[width=0.25\textwidth]{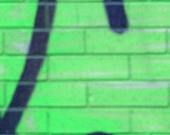} 
 \includegraphics[width=0.25\textwidth]{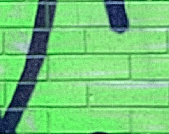}
 \caption{An illustration of unsharp masking. Left: the input image. Right: the resulting image. The unsharp masking procedure helps preserving the edge information of the input image to ensure that the vertices in unclear edges can also be extracted.}
 \label{fig:unsharp}
\end{figure}

Here, $*$ is the convolution operator and $M_s$ is the disk with center $(x,y)$ and radius $s$. The effect of the unsharp masking is demonstrated in Figure \ref{fig:unsharp}. With the aid of this step, we can highlight the edge information in the resulting image $\bar{I}$ for the construction of the triangulation in the later steps. For simplicity we set $s=\sigma$. In our experiment, we choose $\lambda = 0.5,~\sigma = 2,~s = 2$, and $\theta = 0.5$. An analysis on the choice of the parameters is provided in Section \ref{experiment}.

\subsubsection{Segmenting the image}
After obtaining the image $\bar{I}$ upon unsharp masking, we perform a segmentation in this step in order to optimally locate the mesh vertices for computing the coarse triangulation. Mathematically, our segmentation problem is described as follows.

Suppose the image $\bar{I}$ has $L$ intensity levels in each RGB channel. Denote $i$ as a specific intensity level $(i.e. ~ 0 \leq i \leq L-1)$. Let $C$ be a color channel of the image $(i.e. ~ C \in \lbrace R,G,B \rbrace)$, and let $h_{i}^{C}$ denote the image histogram for channel $C$, in other words, the number of pixels which correspond to its $i$-th intensity level.

Define $p_{i}^{C} \triangleq \frac{h_{i}^{C}}{N}$, where $N$ represents the total number of pixels in the image $\bar{I}$. Then we have
\begin{equation}
\sum\limits_{\substack{i = 0, \\  C \in \lbrace R,G,B \rbrace}}^L p_{i}^{C} = 1 \ \ \text{ and } \ \ \mu_{T}^{C} = \sum\limits_{\substack{i = 0, \\  C \in \lbrace R,G,B \rbrace}}^L ip_{i}^{C}.
\end{equation}

Suppose that we want to compress the color space of the image $\bar{I}$ to $l$ intensity levels. Equivalently, $\bar{I}$ is to be segmented into $l$ classes $D_{1}^{C},\cdots,D_{l}^{C}$ by the ordered threshold levels $ x_{j}^{C}, j = 1,\cdots,l-1$. We define the best segmentation criterion to be maximizing the inter-class intensity-mean variance. More explicitly, we define the cost
\begin{equation}\label{eqt:pso_cost}
 \sigma^C \triangleq \sum\limits_{\substack{j = 1, \\  C \in \lbrace R,G,B \rbrace}}^l w_{j}^{C}(\mu_{j}^{C} - \mu_{T}^{C})^2,
\end{equation}
where the probability $w_{j}^{C}$ of occurrence of a pixel being in the class $D_j^C$ and the intensity-mean $\mu_{j}^{C}$ of the class $D_{j}^{C}$ are respectively given by
\begin{equation}
 w_{j}^{C} =
\begin{cases}
\sum\limits_{\substack{i = 0, \\  C \in \lbrace R,G,B \rbrace}}^{t_{j}^{C}} p_{i}^{C} & \text{ if } j = 1, \\
\sum\limits_{\substack{i = t_{j-1}^{C} + 1, \\  C \in \lbrace R,G,B \rbrace}}^{t_{j}^{C}} p_{i}^{C} & \text{ if } 1 < j < l, \\
\sum\limits_{\substack{i = t_{j}^{C} + 1, \\  C \in \lbrace R,G,B \rbrace}}^{L - 1} p_{i}^{C} & \text{ if } j = l,
\end{cases}
 \ \text{ and } \
\mu_{j}^{C} =
\begin{cases}
\sum\limits_{\substack{i = 0, \\  C \in \lbrace R,G,B \rbrace}}^{t_{j}^{C}} \frac{ip_{i}^{C}}{w_{j}^{C}} & \text{ if } j = 1, \\
\sum\limits_{\substack{i = t_{j-1}^{C} + 1, \\  C \in \lbrace R,G,B \rbrace}}^{t_{j}^{C}} \frac{ip_{i}^{C}}{w_{j}^{C}}
& \text{ if } 1 < j < l, \\
\sum\limits_{\substack{i = t_{j}^{C} + 1, \\  C \in \lbrace R,G,B \rbrace}}^{L - 1} \frac{ip_{i}^{C}}{w_{j}^{C}}
& \text{ if }  j = l.
\end{cases}
\end{equation}

Hence, we maximize three objective functions of each RGB channel
\begin{equation}
\underset{1 < x_{1}^{C} < \cdots < x_{l-1}^{C} < L}{\arg\max} \sigma^C(\lbrace x_{j}^{C} \rbrace_{j = 1}^{l-1}),
\end{equation}
where $C \in \{R, G, B\}$. Our goal is to find a set of $\textbf{x} = \lbrace x_{j}^{C} \rbrace_{j = 1}^{l-1}$ such that above function is maximized for each RGB channel.

To solve the aforementioned segmentation optimization problem, we apply the Particle Swarm Optimization (PSO) segmentation algorithm \cite{Ghamisi12} on the image $\bar{I}$. The PSO method is used in this segmentation optimization problem for reducing the chance of trapping in local optimums.

\begin{figure}[t]
 \centering
 \includegraphics[width=0.25\textwidth]{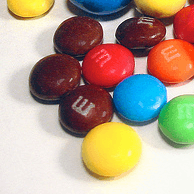} 
 \includegraphics[width=0.25\textwidth]{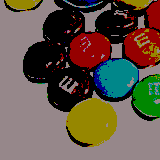}
 \caption{An illustration of the segmentation step for compressing the color space to achieve a sparse intensity representation. Left: the original image. Right: the segmentation result.}
 \label{fig:pso}
\end{figure}

An illustration of the segmentation step is provided in Figure \ref{fig:pso}. After performing the segmentation, we extract the boundaries of the segments. Then, we can obtain a number of large patches of area in each of which the intensity information is almost the same. They provide us with a reasonable edge base for constructing a coarse triangulation in later steps.

\subsubsection{Sparse feature extraction on the segment boundaries}
After computing the segment boundaries $\mathcal{B}$ on the image $\bar{I}$, we aim to extract sparse feature points on $\mathcal{B}$ in this step. For the final triangulation, it is desirable that the edges of the triangles are as close as possible to the segment boundaries $\mathcal{B}$, so as to preserve the geometric features of the original image $I$. Also, to improve the efficiency for the computations on the triangulation, the triangulation should be much coarser than the original image. To achieve the mentioned goals, we consider extracting sparse features on the segment boundaries $\mathcal{B}$ and use them as the vertices of the ultimate triangulated mesh.

Consider a rectangular grid table $G$ on the image $\bar{I}$. Apparently, the grid table $G$ intersects the segment boundaries $\mathcal{B}$ at a number of points. Denote $\mathcal{P}$ as our desired set of sparse features. Conceptually, $\mathcal{P}$ is made up of the set of points at which $\mathcal{B}$ intersect the grid $G$, with certain exceptions.

\begin{figure}[t]
 \centering
 \includegraphics[width=0.6\textwidth]{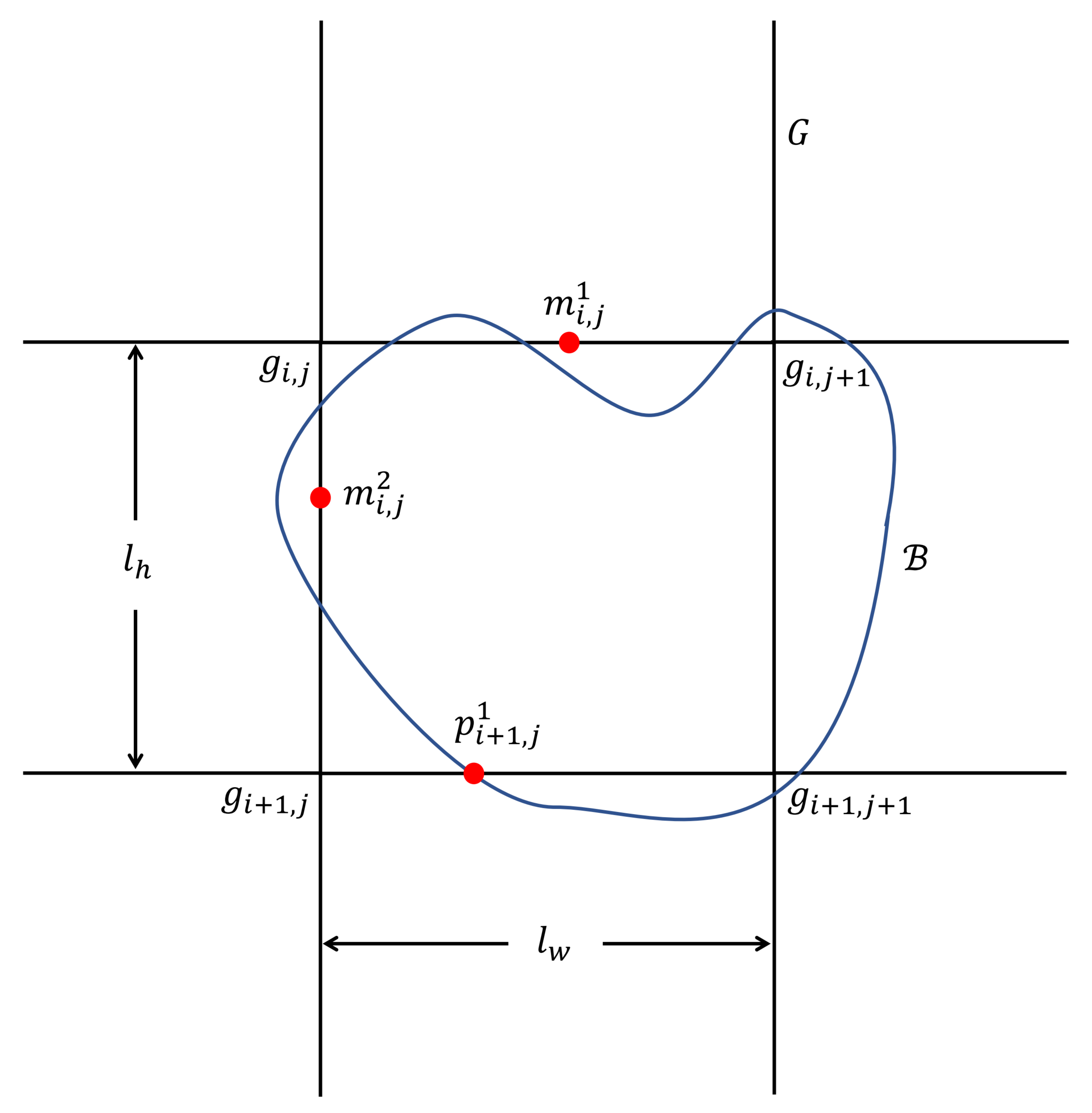}
 \caption{An illustration of our sparse feature extraction scheme. The chosen sparse feature points are represented by the red dots. If the segment boundary does not intersect an edge, no point is selected. If the segment boundary intersects an edge at exactly one point, the point is selected as a feature point. If the segment boundary intersects an edge at multiple points, the centre of the points is selected as a feature point.}
 \label{fig:feature}
\end{figure}

In order to further reduce the number of feature points for a coarse triangulation, we propose a merging procedure for close points. Specifically, let $g_{i,j}$ be the vertex of the grid $G$ at the $i$-th row and the $j$-th column. We denote $\mathcal{P}_{i,j}^1$ and $\mathcal{P}_{i,j}^2$ respectively as the set of points at which $\mathcal{B}$ intersect the line segment $\displaystyle \overline{g_{i,j} g_{i,j+1}}$ and the line segment $\displaystyle \overline{g_{i,j} g_{i+1,j}}$. See Figure \ref{fig:feature} for an illustration of the parameters.

There are 3 possible cases for $\mathcal{P}_{i,j}^k$, where $k = 1,2$:
\begin{enumerate}[(i)]
 \item If $|\mathcal{P}_{i,j}^k|=0$, then there is no intersection point between the line segment and $\mathcal{B}$ and hence we can neglect it.
 \item If $|\mathcal{P}_{i,j}^k|=1$, then there is exactly one intersection point $p_{i,j}^k$ between the line segment and $\mathcal{B}$. We include this intersection point $p_{i,j}^k$ in our desired set of sparse features $\mathcal{P}$.
 \item If $|\mathcal{P}_{i,j}^k|>1$, then there are multiple intersection points between the line segment and $\mathcal{B}$. Since these multiple intersection points lie on the same line segment, it implies that they are sufficiently close to each other. In other words, the information they contain about the segment boundaries $\mathcal{B}$ is highly similar and redundant. Therefore, we consider merging these multiple points as one point.
\end{enumerate}
More explicitly, for the third case, we compute the centre $m_{i,j}^k$ of the points in $\mathcal{P}_{i,j}^k$ by
\begin{equation}
\displaystyle m_{i,j}^k = mean_{\{p | p \in \mathcal{P}_{i,j}^k\}} p.
\end{equation}
The merged point $m_{i,j}^k$ is then considered as a desired feature point. In summary, our desired set of sparse features is given by
\begin{equation}
\displaystyle \mathcal{P} = \bigcup_i \bigcup_j \{p_{i,j}^1, p_{i,j}^2, m_{i,j}^1, m_{i,j}^2\}.
\end{equation}

An illustration of the sparse feature extraction scheme is given in Figure \ref{fig:feature}.
However, one important problem in this scheme is to determine a suitable size of the grid $G$ so that the sparse feature points are optimally computed. Note that to preserve the regularity of the extracted sparse features, it is desirable that the elements of the grid $G$ are close to perfect squares. Also, to capture the important features as complete as possible, the elements of $G$ should be small enough. Mathematically, the problem can be formulated as follows.

Denote $w$ as the width of the image $\bar{I}$, $h$ as the height of the image $\bar{I}$, $w'$ as the number of columns in $G$, $h'$ as the number of rows in $G$, $l_{w}$ as the horizontal length of every element of $G$, and $l_{h}$ as the vertical length of every element of $G$. See Figure \ref{fig:feature} for a geometric illustration of $l_w$ and $l_h$. We further denote $p$ as the percentage of grid edges in $G$ which intersect the segment boundaries $\mathcal{B}$, and $n$ as the desired number of the sparse feature points. Given the two inputs $p$ and $n$, to find a suitable grid size of $G$, we aim to minimize the cost function
\begin{equation}
c(l_{w},l_{h}) = \left| l_{w} - l_{h} \right|^2
\end{equation}
subject to
\begin{align}
  \mbox{(i)} & \qquad h = h'l_{h},\\
 \mbox{(ii)} &  \qquad w = w'l_{w},\\
 \mbox{(iii)} &  \qquad p(w'+h'+2w'h') = n.\label{eqt:number}
\end{align}

Here, the first and the second constraint respectively correspond to the horizontal and vertical dimensions of the grid $G$, and the third constraint corresponds to the total number of intersection points. To justify Equation (\ref{eqt:number}), note that

\begin{equation}
\begin{split}
&\text{Total \# of line segments} \\
= \ &\text{Total \# of horizontal line segments} + \text{Total \# of vertical line segments}\\
= \ &h'(w'+1) + w'(h'+1)\\
= \ &w'+h'+2w'h'.\\
\end{split}
\end{equation}

Note that this minimization problem is nonlinear. To simplify the computation, we assume that $w'$, $h'$ are very large, that is, the grid $G$ is sufficiently dense. Then, from Equation (\ref{eqt:number}), we have
\begin{equation}
\begin{split}
\frac{p}{n} &= \frac{1}{w'+h'+2w'h'} \approx \frac{1}{2w'h'} = \frac{1}{2\left(\frac{w}{l_w}\right) \left(\frac{h}{l_h}\right)} = \frac{l_{w}l_{h}}{2wh}.
\end{split}
\end{equation}
By further assuming that the grid $G$ is sufficiently close to a square grid, we have $l_{w} \approx l_{h}$. Then, it follows that
\begin{equation}
\frac{p}{n} \approx \frac{l_{w}^2}{2wh},\quad
l_{w} \approx \sqrt{\frac{2pwh}{n}}.
\end{equation}
Similarly,
\begin{equation}
l_{h} \approx \sqrt{\frac{2pwh}{n}}.
\end{equation}

To satisfy the integral constraints for $w'$ and $h'$, we make use of the above approximations and set
\begin{equation}
h' = h_0' := \left\lfloor \frac{h}{\sqrt{\frac{2pwh}{n}}} \right\rfloor = \left\lfloor \sqrt{\frac{nh}{2pw}} \right\rfloor.
\end{equation}
Similarly, we set
\begin{equation}
w' = w_0' := \left\lfloor \frac{w}{\sqrt{\frac{2pwh}{n}}}  \right\rfloor = \left\lfloor \sqrt{\frac{nw}{2ph}} \right\rfloor.
\end{equation}
Finally, we take
\begin{equation}
l_{h} = \frac{h}{h_0'}
 \ \text{ and } \
l_{w} = \frac{w}{w_0'}.
 \end{equation}

To summarize, with the abovementioned strategy for the feature point extraction, we obtain a set of sparse feature points which approximates the segment boundaries $\mathcal{B}$. Specifically, given the inputs $p$ and $n$, the rectangular grid $G$ we introduce leads to approximately $n$ regularly-extracted sparse feature points. An illustration of the sparse feature extraction scheme is shown in Figure \ref{fig:delaunay} (left). In our experiments, $p$ is set to be 0.2, and $n$ is set to be $10\%$ of the number of pixels in the segmentation result. A denser triangulated representation can be achieved by increasing the value of $p$.


\subsubsection{Adding landmark points to the vertex set of the desired coarse triangulation}
This step is only required when our {\em TRIM} algorithm is used for landmark-constrained registration. For accurate landmark-constrained registration, it is desirable to include the landmark points in the vertex set of the coarse representations of the input image $I$. One of the most important features of our coarse triangulation approach is that it allows registration with exact landmark constraints on a coarse triangular representation. By contrast, the regular grid-based registration can only be achieved on very dense rectangular grid domains in order to reduce the numerical errors.

With the abovementioned advantage of our approach, we can freely add a set of landmark points $\mathcal{P}_{LM}$ to the set of sparse features $\mathcal{P}$ extracted by the previous procedure. In other words, the landmark points are now considered as a part of our coarse triangulation vertices:
\begin{equation}
\mathcal{P} = \bigcup_i \bigcup_j \{p_{i,j}^1, p_{i,j}^2, m_{i,j}^1, m_{i,j}^2\} \cup \mathcal{P}_{LM}.
\end{equation}
Then, the landmark-constrained registration of images can be computed by the existing feature-matching techniques for triangular meshes. The existing feature detection approaches such as \cite{harris88} and \cite{Lowe04} can be applied for obtaining the landmark points.

\subsubsection{Computing a Delaunay triangulation}
In the final step, we construct a triangulation based on the set $\mathcal{P}$ of feature points. Among all triangulation schemes, the Delaunay triangulation method is chosen since the triangles created by the Delaunay triangulations are more regular. More specifically, if $\alpha$ and $\beta$ are two angles opposite to a common edge in a Delaunay triangulation, then they must satisfy the inequality
\begin{equation}
 \alpha + \beta \leq \pi.
\end{equation}

\begin{figure}[t]
 \centering
 \includegraphics[width=0.27\textwidth]{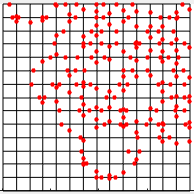}
 \includegraphics[width=0.27\textwidth]{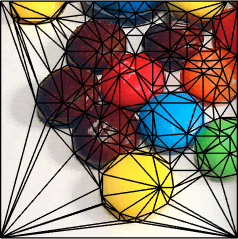}
 \includegraphics[width=0.27\textwidth]{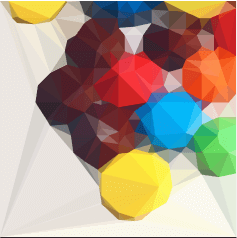}
 \caption{An illustration of computing a Delaunay triangulation on the extracted features. Left: the points obtained by the feature extraction step from Figure \ref{fig:pso}. Middle: a Delaunay triangulation on the feature points. Right: the triangulation with a color approximated on each triangle.}
 \label{fig:delaunay}
\end{figure}

In other words, Delaunay triangulations always aim to minimize the formation of sharp and irregular triangles. Note that the regularity does not only enhance the visual quality of the resulting triangulation but also lead to a more stable approximation of the derivatives on the triangles when applying various registration schemes. Therefore, we compute a Delaunay triangulation  on the set $\mathcal{P}$ of feature points for achieving the ultimate triangulation $\mathcal{T}$. An illustration of the construction of the Delaunay triangulations is shown in Figure \ref{fig:delaunay}.

These 6 steps complete our {\em TRIM} algorithm as summarized in Algorithm \ref{algorithm:triangulation}.
\begin{algorithm}[h]
\footnotesize
\KwIn{An image $I$, the desired number of image intensity levels $l$ for segmentation, the desired number of feature points $n$, the sparse ratio $p$.}
\KwOut{A coarse triangulation $\mathcal{T}$ that captures the main features of the image.}
\BlankLine
 Subsample the input image $I$ to a suitable size and denote the result by $\tilde{I}$\;
 Apply an unsharp masking on the subsampled image $\tilde{I}$ and denote the result by $\bar{I}$\;
 Apply the PSO segmentation for compressing the color space of $\bar{I}$ to $l$ intensity levels, and extract boundaries $\mathcal{B}$ of the segments\;
 Extract a set of sparse feature points $\mathcal{P}$ from the segment boundaries $\mathcal{B}$ based on the parameters $n$ and $p$\;
 Add a set of extra landmark points $\mathcal{P}_{LM}$ to $\mathcal{P}$ if necessary\;
 Compute a Delaunay triangulation $\mathcal{T}$ on the sparse feature points $\mathcal{P}$.
\caption{Our proposed {\em TRIM} algorithm for triangulating images}
\label{algorithm:triangulation}
\end{algorithm}

It is noteworthy that our proposed {\em TRIM} algorithm significantly trims down high resolution images without distorting their important geometric features. Experimental results are shown in Section \ref{experiment} to demonstrate the effectiveness of the {\em TRIM} algorithm.

\subsection{Stage $2$ -- Registration of two triangulated image surfaces}
With the above triangulation algorithm for images, we can simplify the image registration problem as a mapping problem of triangulated surfaces rather than of sets of landmark points. Many conventional image registration approaches are hindered by the long computational time and the accuracy of the initial maps. With the new strategy, it is easy to obtain a highly efficient and reasonably accurate registration of images. Our registration result can serve as a high quality initial map for various algorithms.

\begin{figure}[t]
 \centering
 \includegraphics[width=0.6\textwidth]{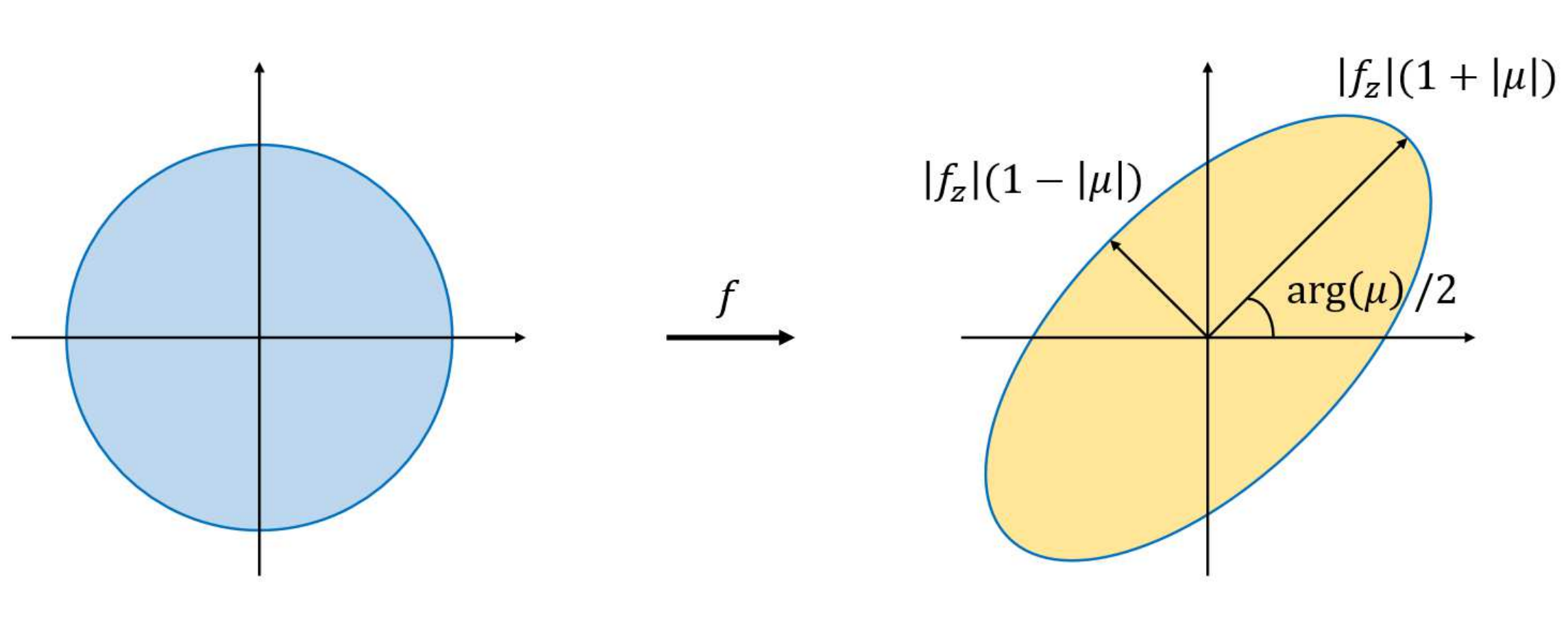}
 \caption{An illustration of quasi-conformal mappings. The maximal magnification and shrinkage are determined by the Beltrami coefficient $\mu$ of the mappings.}
 \label{fig:beltrami}
\end{figure}

To preserve angles and hence the local geometry of two surfaces, rather than simply mapping two sets of points, conformal mappings may not exist due to presence of landmark constraints. We turn to consider quasi-conformal mappings, a type of mappings which is closely related to the conformal mappings. Mathematically, a \emph{quasi-conformal mapping} $f: \mathbb{C} \to \mathbb{C}$ satisfies the Beltrami equation
\begin{equation}
\frac{\partial f}{\partial \bar{z}} = \mu(z) \frac{\partial f}{\partial z}
\end{equation}
where $\mu$ (called the \emph{Beltrami coefficient} of $f$) is a complex-valued function with sup norm less than 1. Intuitively, a conformal mapping maps infinitesimal circles to infinitesimal circles, while a quasi-conformal mapping maps infinitesimal circles to infinitesimal ellipses (see Figure \ref{fig:beltrami}). Readers are referred to \cite{Gardiner00} for more details.

In this work, we apply the quasi-conformal landmark registration (QCLR) algorithm (designed for general surfaces in \cite{Lam14}) to our coarse triangulations of images. More explicitly, to compute a registration mapping $f:I_1 \to I_2$ between two images $I_1$ and $I_2$ with prescribed point correspondences
\begin{equation} \label{eqt:constraint}
p_i \longleftrightarrow q_i, i = 1,2,\cdots,n,
\end{equation}
where $\{p_i\}_{i=1}^n$ are a set of points on $I_1$ and $\{q_i\}_{i=1}^n$ are a set of points on $I_2$, we first apply our proposed {\em TRIM} algorithm and obtain a coarse triangulation $\mathcal{T}_1$ on $I_1$. Here, we include the feature points $\{p_i\}_{i=1}^n$ in the generation of the coarse triangulation, as described in the fifth step of the {\em TRIM} algorithm. Then, instead of directly computing $f$, we can solve for a map $\tilde{f}: \mathcal{T}_1 \to I_2$. Since the problem size is significantly reduced under the coarse triangulation, the computation for $\tilde{f}$ is much more efficient than that for $f$.

The QCLR algorithm makes use of the penalty splitting method and minimizes
\begin{equation}
 E_{LM}^{split}(\nu,\tilde{f}) = \int_{\mathcal{T}_1} |\nabla \nu|^2 + \alpha \int_{\mathcal{T}_1} |\nu|^2 + \rho \int_{\mathcal{T}_1} |\nu - \mu(\tilde{f})|^2
\end{equation}
subject to
(i) $\tilde{f}(p_i) = q_i$ for all $i = 1,2,\cdots,n$ and
(ii) $\|\nu\|_{\infty} <1$.
Further alternating minimization of the energy $E_{LM}^{split}$ over $\nu$ and $\tilde{f}$ is used. Specifically, for computing $\tilde{f}_n$ while fixing $\nu_n$ and the landmark constraints, we apply the linear Beltrami solver by Lui et al.\cite{Lui13}. For computing $\nu_{n+1}$ while fixing $\tilde{f}_n$,
by considering the Euler-Lagrange equation, it suffices to solve
\begin{equation}
 (-\Delta + 2 \alpha I + 2\rho I) \nu_{n+1} = 2\rho \mu(\tilde{f}_n).
\end{equation}
From $\nu_{n+1}$, one can compute the associated quasi-conformal mapping $\tilde{f}_{n+1}$ and then update  $\nu_{n+1}$ by
\begin{equation}
 \nu_{n+1} \leftarrow \nu_{n+1} + t(\mu(\tilde{f}_{n+1}) - \nu_{n+1})
\end{equation}
for some small $t$ to satisfy the landmark constraints (\ref{eqt:constraint}).

After computing the quasi-conformal mapping $\tilde{f}$ on the coarse triangulation, we interpolate once to retrieve the fine details of the registration in the high resolution. Since the triangulations created by our proposed {\em TRIM} algorithm preserves the important geometric features and prominent straight lines of the input image, the details of the registration results can be accurately interpolated. Moreover, since the coarse triangulation largely simplifies the input image and reduces the problem size, the computation is significantly accelerated.

The overall registration procedure is summarized in Algorithm \ref{algorithm:registration}. Experimental results are illustrated in Section \ref{experiment} to demonstrate the significance of our coarse triangulation in the registration scheme.

\begin{algorithm}[h]
\footnotesize
\KwIn{Two images or video frames $I_1$, $I_2$ to be registered, with the prescribed feature correspondences.}
\KwOut{A feature-matching registration mapping $f:I_1 \to I_2$.}
\BlankLine
Compute a coarse triangulation $\mathcal{T}_1$ of $I_1$ using our proposed {\em TRIM} algorithm (Algorithm \ref{algorithm:triangulation}). Here, we include the prescribed feature points on $I_1$ in the generation of the coarse triangulation $\mathcal{T}_1$\;
Select landmark correspondences of the coarse triangulation $\mathcal{T}_1$ and the target image $I_2$. Denote the landmark points on $\mathcal{T}_1$ and $I_2$ by $\{p_i\}_{i=1}^n$ and $\{q_i\}_{i=1}^n$ correspondingly\;
Compute a landmark based quasi-conformal mapping $\tilde{f}: \mathcal{T}_1 \to \mathbb{C}$ by the QCLR algorithm in \cite{Lam14}\;
Obtain $f$ by $\tilde{f}$ with a bilinear interpolation between $\mathcal{T}_j$ and $I_j$.
\caption{Feature-based registration via our proposed {\em TRIM} algorithm}
\label{algorithm:registration}
\end{algorithm}

\section{Experimental results} \label{experiment}

In this section, we demonstrate the effectiveness of our proposed triangulation scheme. The algorithms are implemented using MATLAB.  The unsharp masking step is done using MATLAB's \texttt{imsharpen}. The PSO segmentation is done using the MATLAB Central function \texttt{segmentation}. For solving the mentioned linear systems, the backslash operator ($\backslash$) in MATLAB is used. The test images are courtesy of the RetargetMe dataset \cite{Rubinstein10} and the Middlebury Stereo Datasets \cite{Scharstein03,Scharstein07}. The bird image is courtesy of the first author. All experiments are performed on a PC with an Intel(R) Core(TM) i7-4500U CPU @1.80 GHz processor and 8.00 GB RAM.
\begin{figure}[t]
 \centering
 \includegraphics[width=0.21\textwidth]{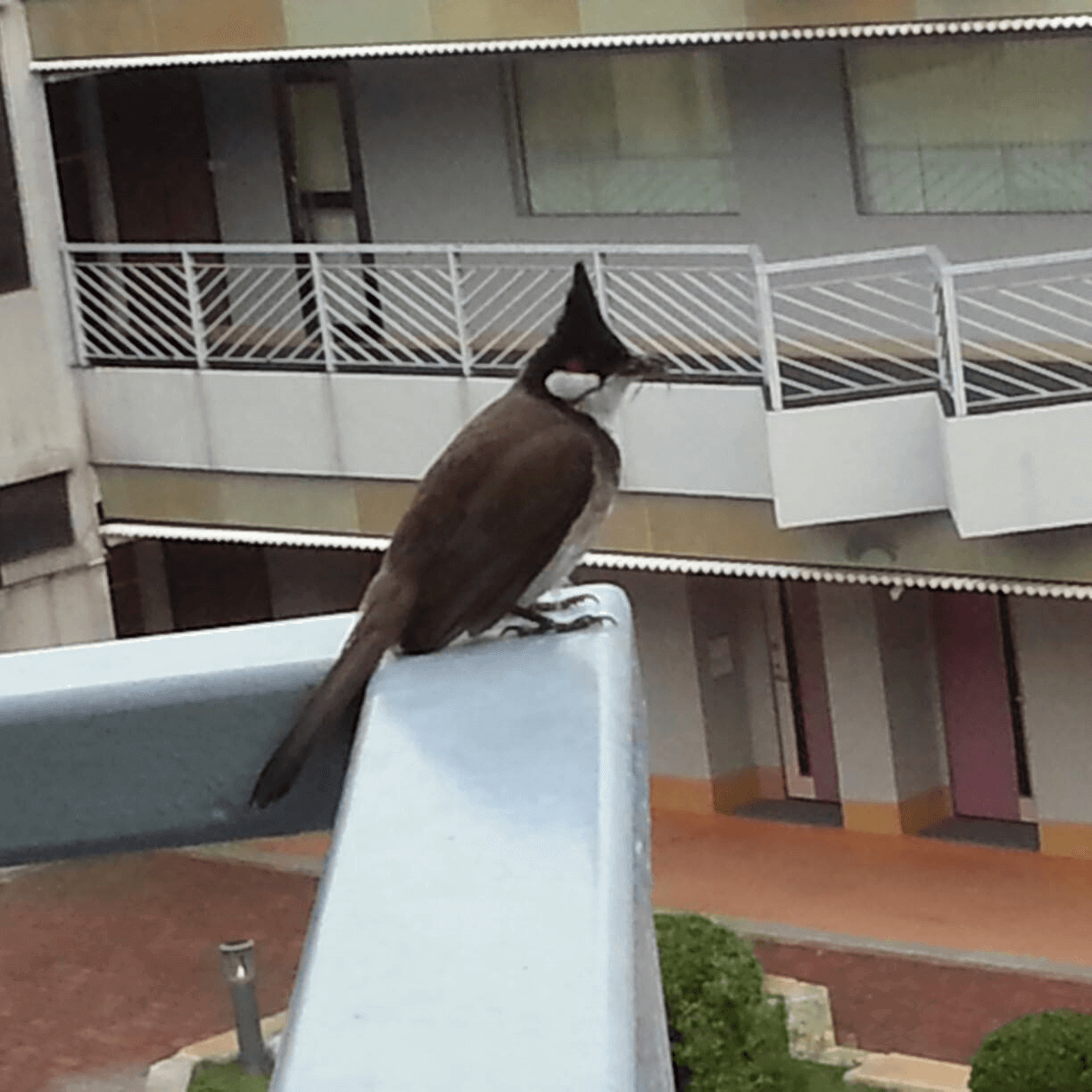} \ 
 \includegraphics[width=0.16\textwidth]{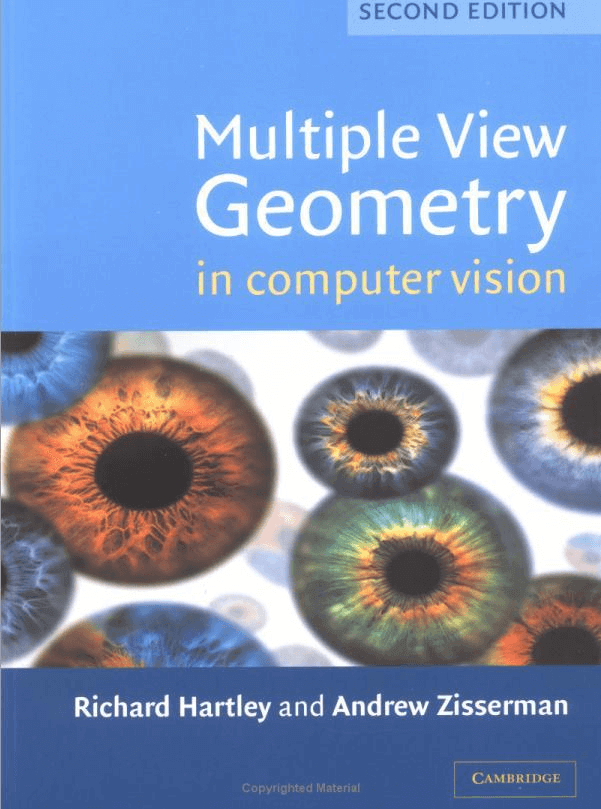} \ 
 \includegraphics[width=0.32\textwidth]{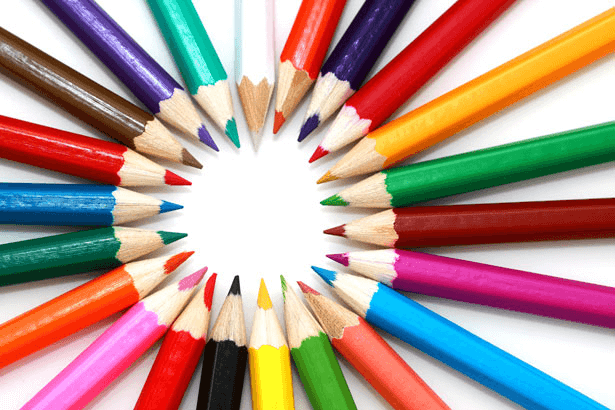} \ 
 \includegraphics[width=0.25\textwidth]{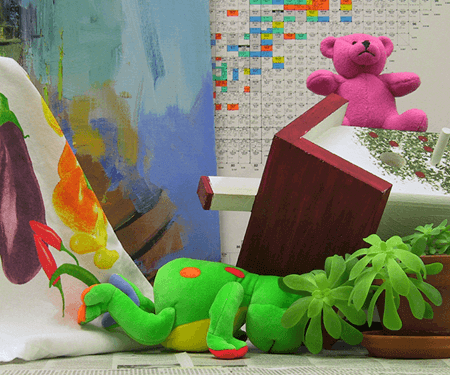}
 \includegraphics[width=0.21\textwidth]{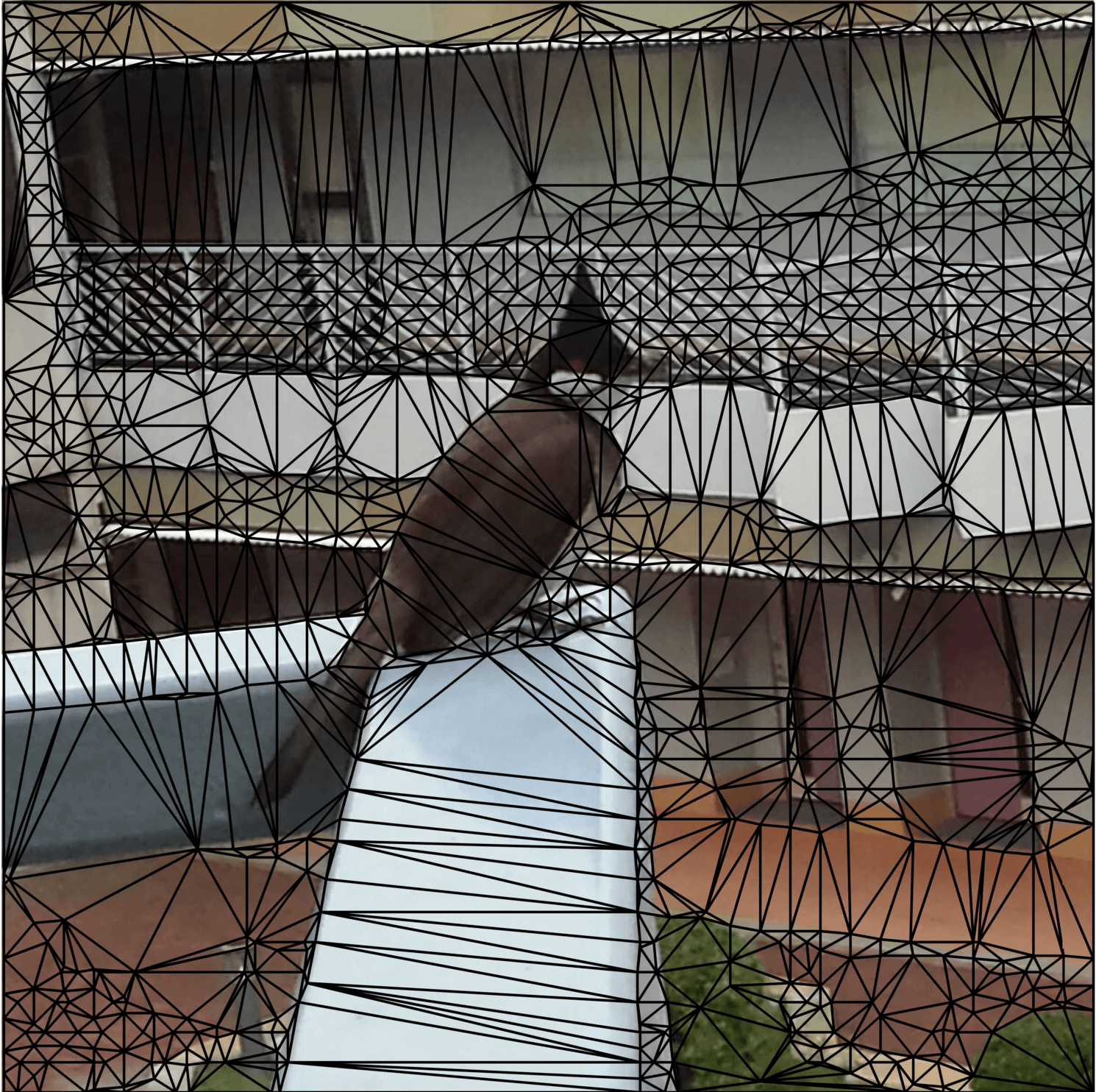} \
 \includegraphics[width=0.16\textwidth]{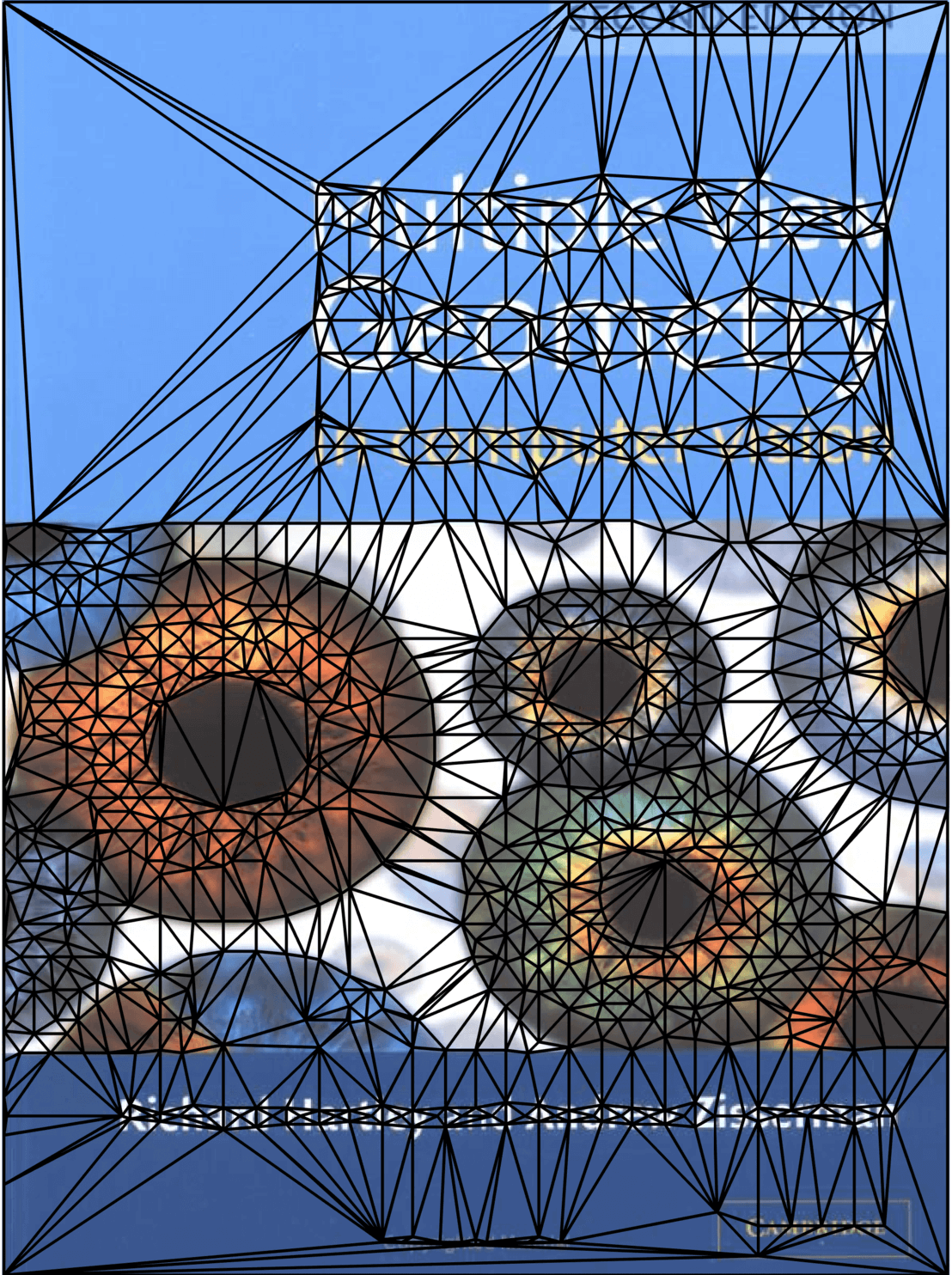} \
 \includegraphics[width=0.32\textwidth]{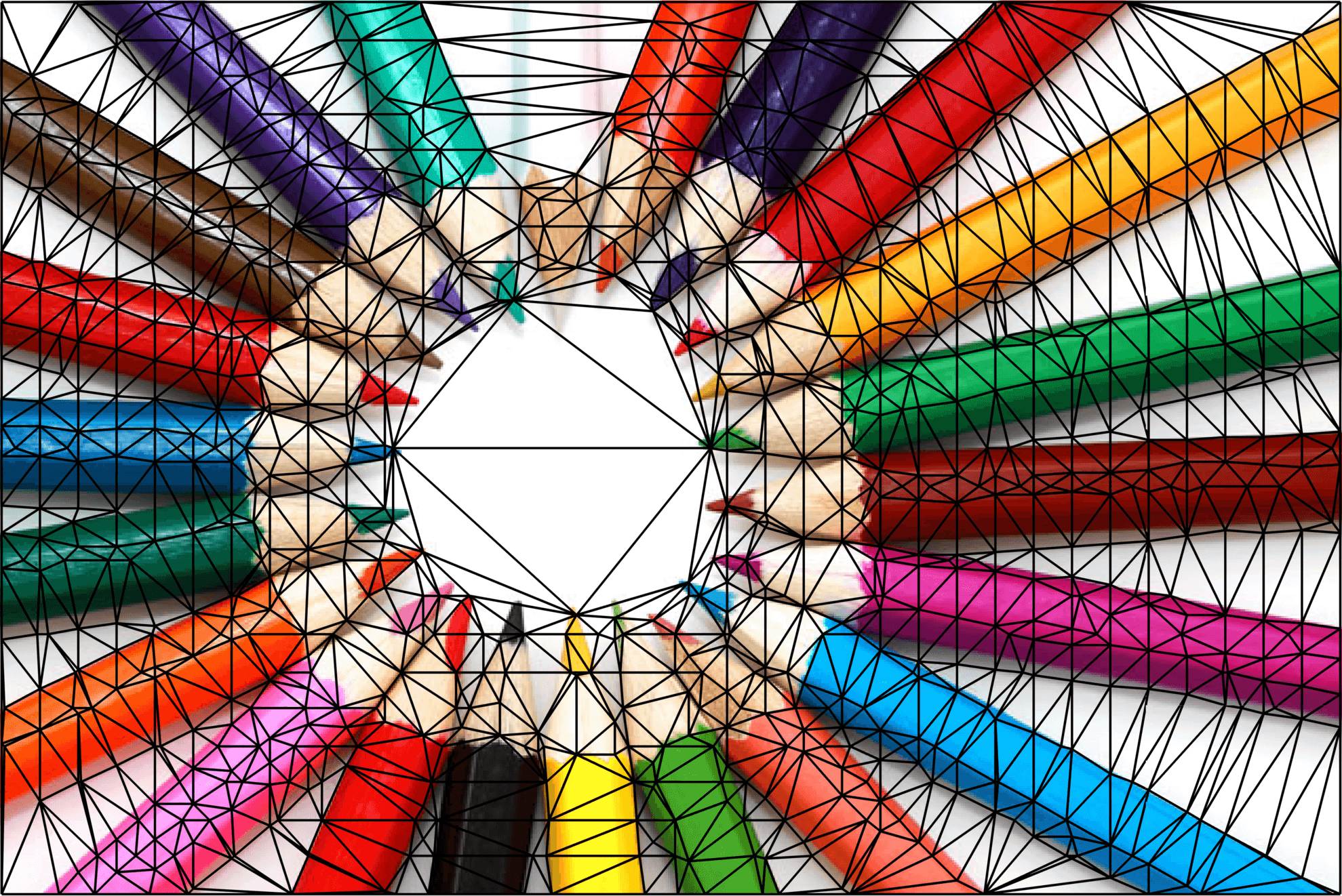} \
 \includegraphics[width=0.25\textwidth]{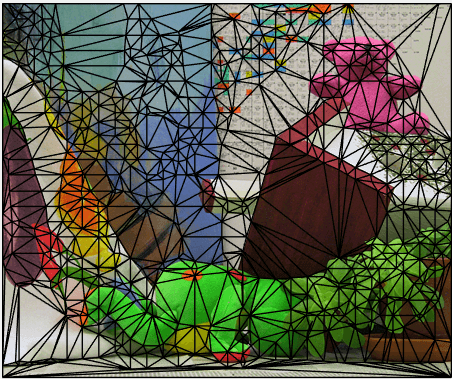}
 \caption{Several images and the triangulations by our {\em TRIM} algorithm. Top: the input images. Bottom: the resulting triangulations. The key features of the images are well represented in our triangulations, and the regions with similar color can be represented by coarse triangulations. }
 \label{fig:triangulation}
\end{figure}

\begin{figure}[t]
 \centering
 \includegraphics[width=0.24\textwidth]{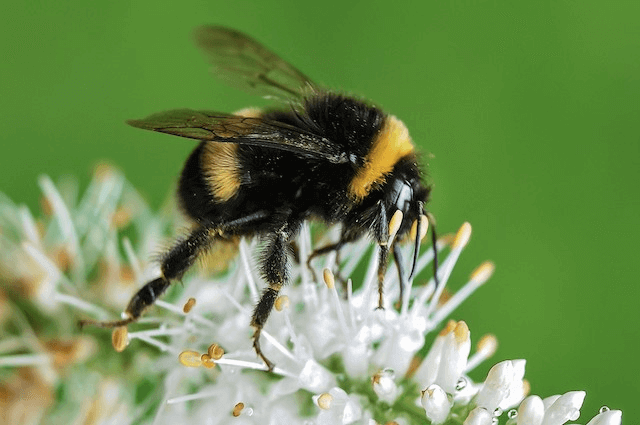}
 \includegraphics[width=0.24\textwidth]{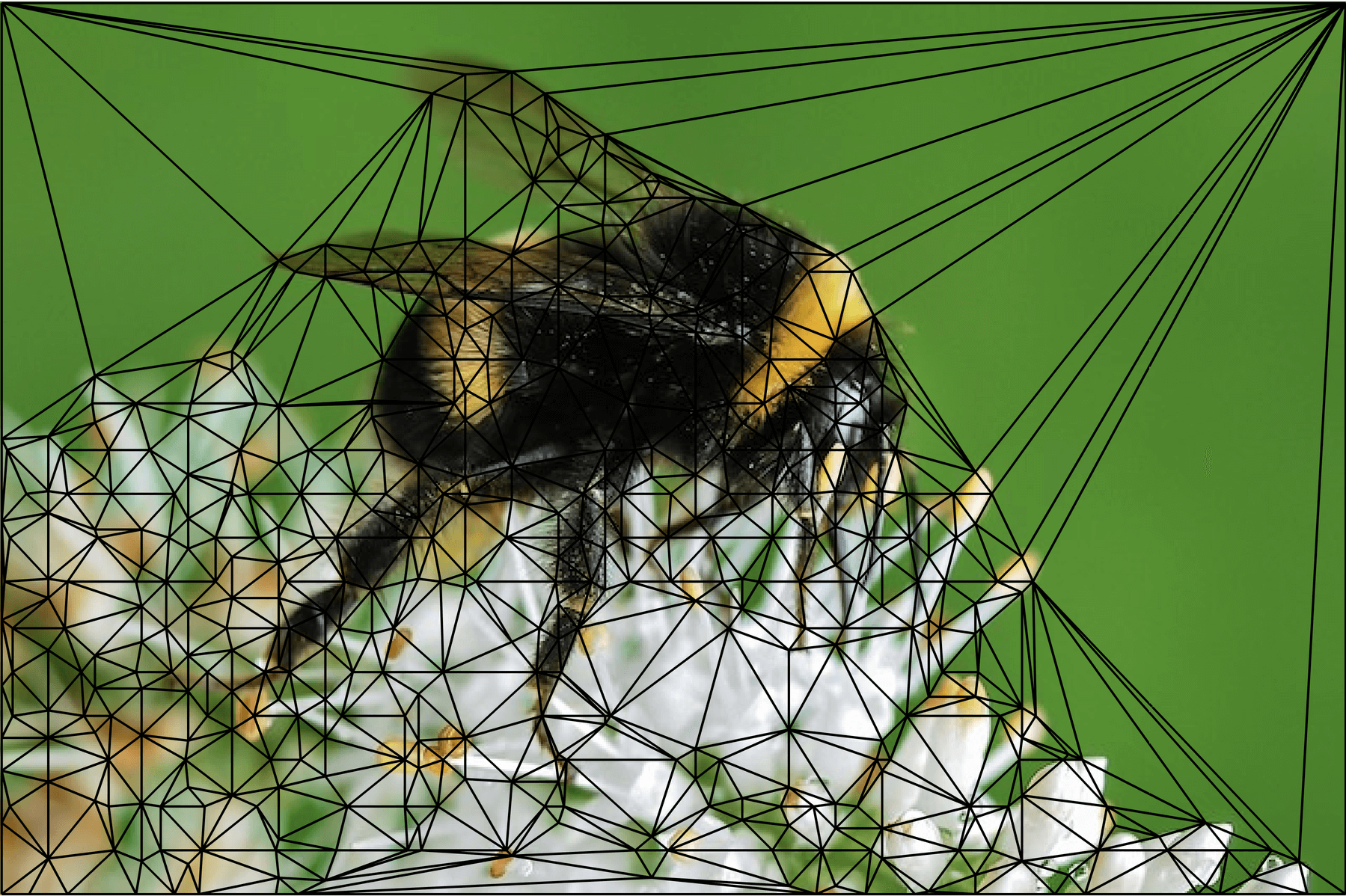}
 \includegraphics[width=0.24\textwidth]{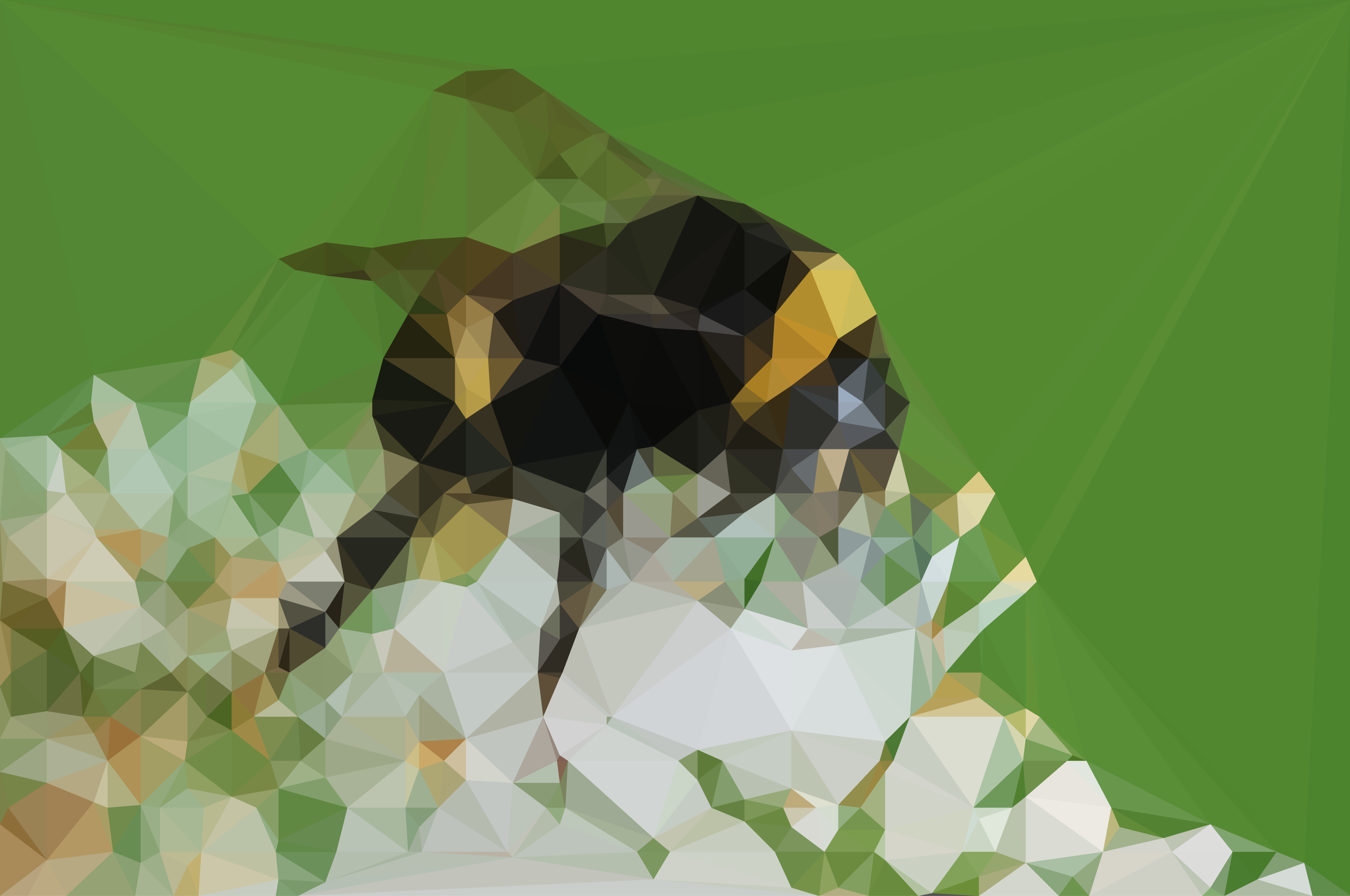}
 \includegraphics[width=0.24\textwidth]{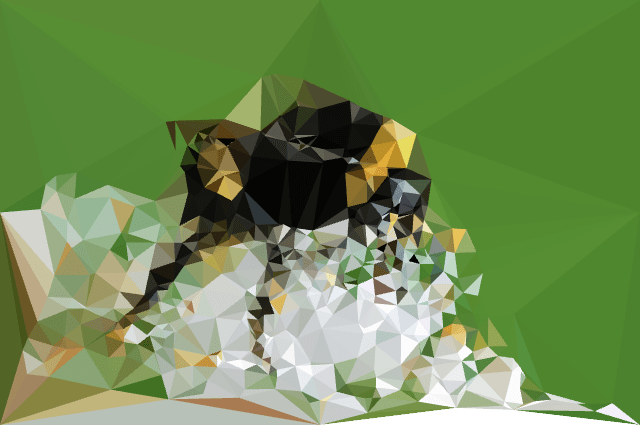}
 \caption{A bee image and the triangulations created by our {\em TRIM} algorithm and DMesh \cite{dmesh}. Left to right: The input image, the coarse triangulation created by {\em TRIM}, our {\em TRIM} coarse triangulation with a color approximated on each triangle, and the triangulation by DMesh \cite{dmesh}.}
 \label{fig:triangulation_bee}
\end{figure}

\begin{figure}[t]
 \centering
 \includegraphics[width=0.24\textwidth]{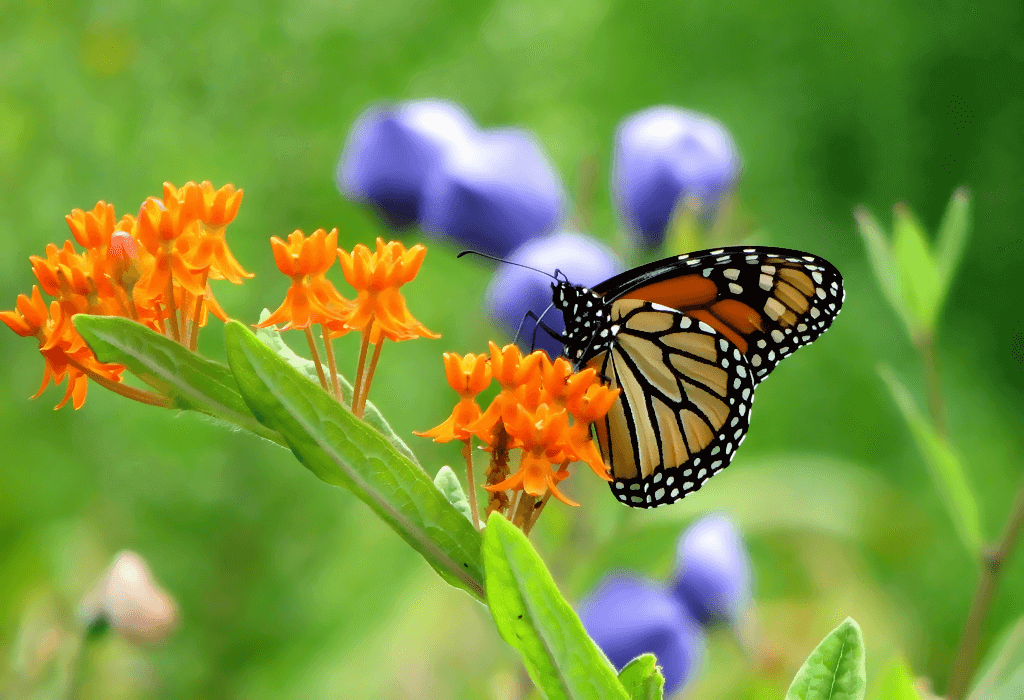}
 \includegraphics[width=0.24\textwidth]{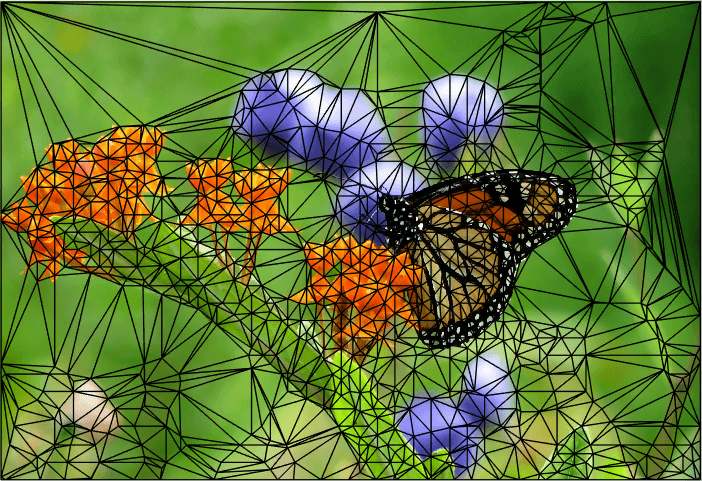}
 \includegraphics[width=0.24\textwidth]{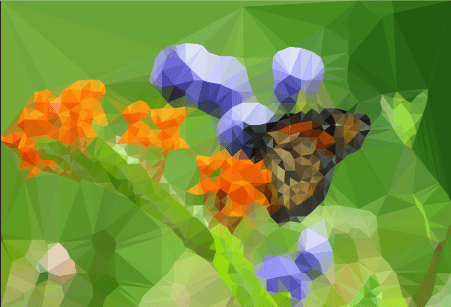}
 \includegraphics[width=0.24\textwidth]{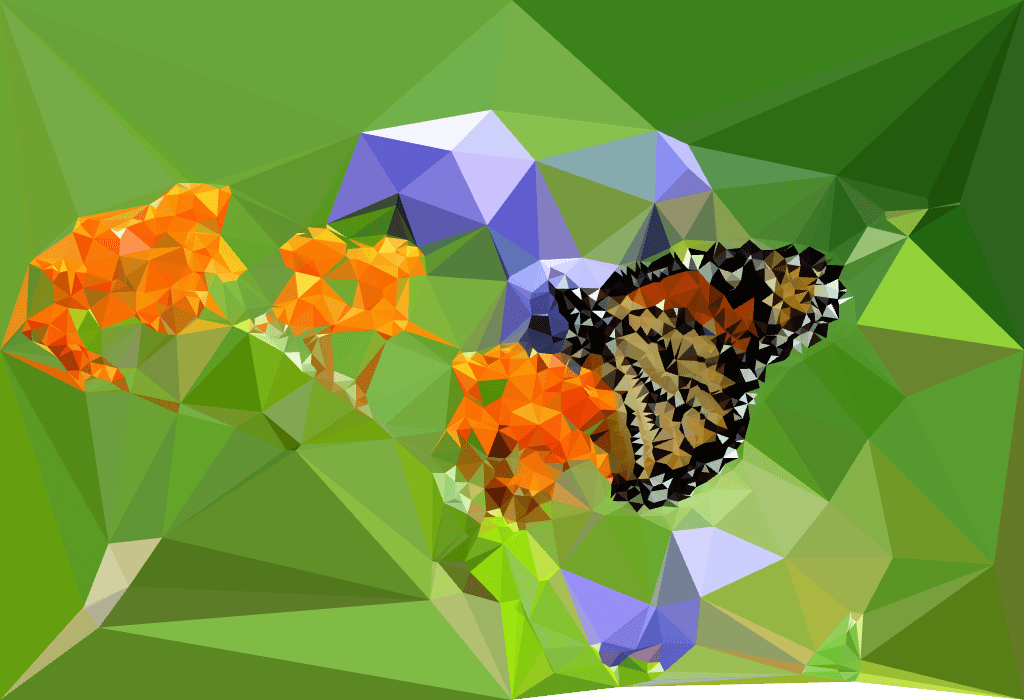}
 \caption{An butterfly image and the triangulations created by our {\em TRIM} algorithm and DMesh \cite{dmesh}. Left to right: The input image, the coarse triangulation created by {\em TRIM}, our {\em TRIM} coarse triangulation with a color approximated on each triangle, and the triangulation by DMesh \cite{dmesh}.}
 \label{fig:triangulation_butterfly}
\end{figure}

\begin{figure}[t]
 \centering
 \includegraphics[width=0.26\textwidth]{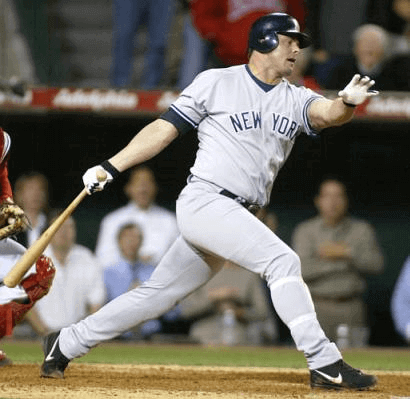}
 \includegraphics[width=0.26\textwidth]{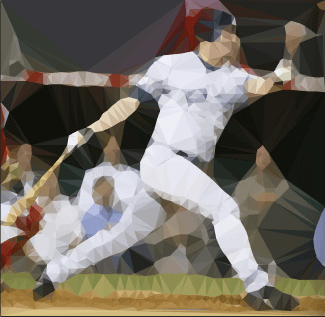}
 \includegraphics[width=0.26\textwidth]{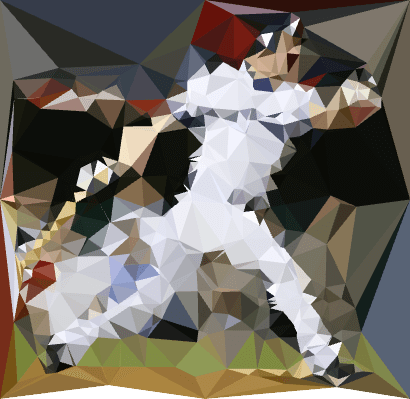}
 \includegraphics[width=0.26\textwidth]{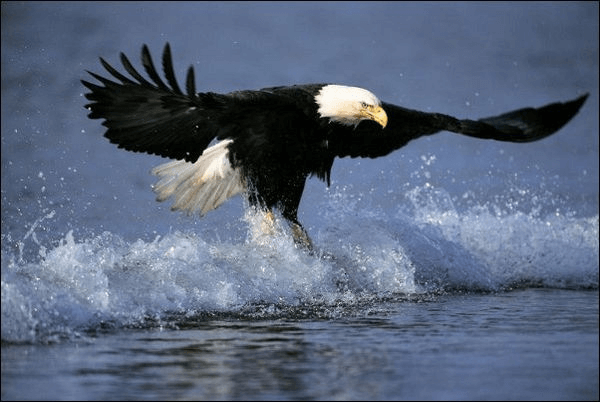}
 \includegraphics[width=0.26\textwidth]{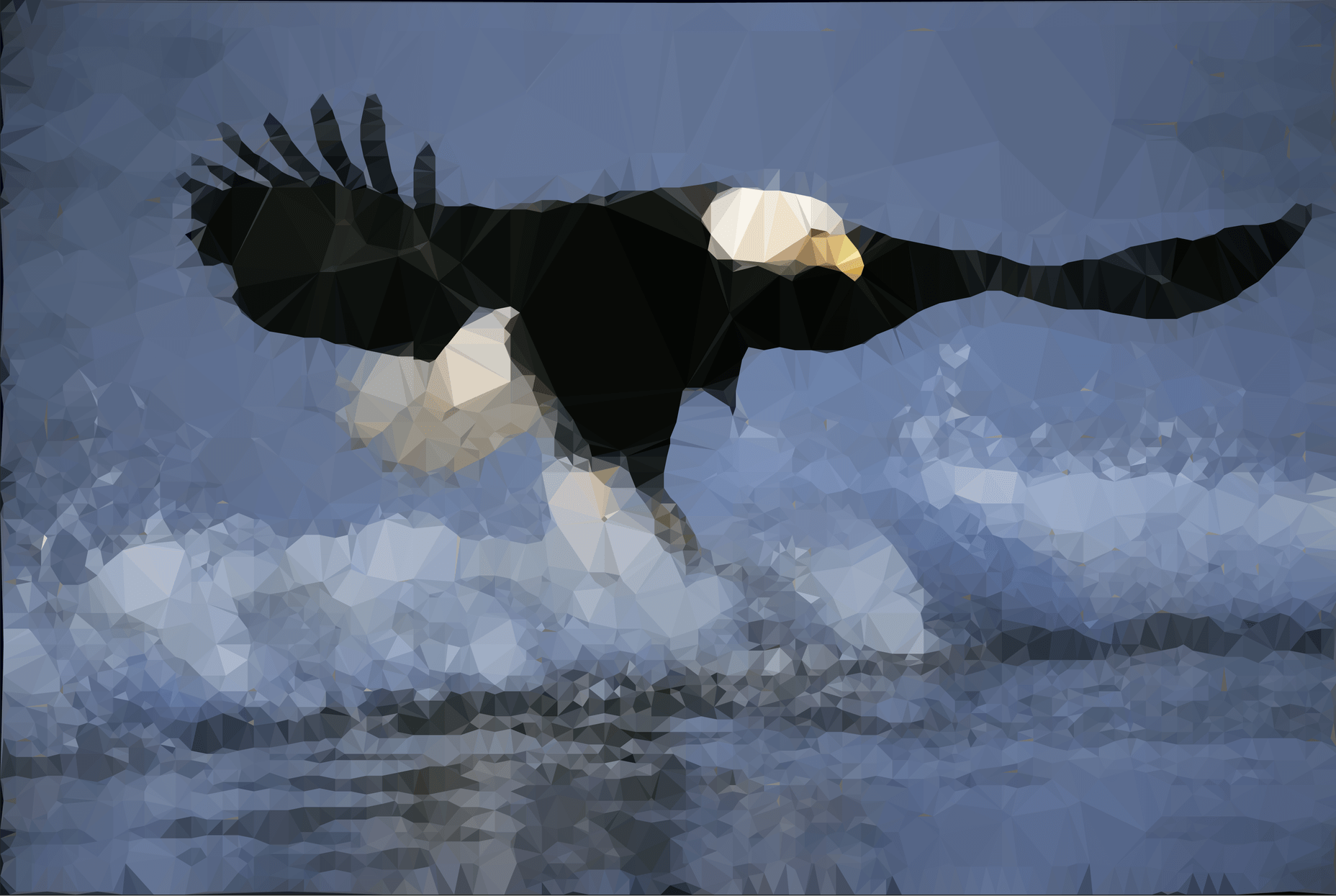}
 \includegraphics[width=0.26\textwidth]{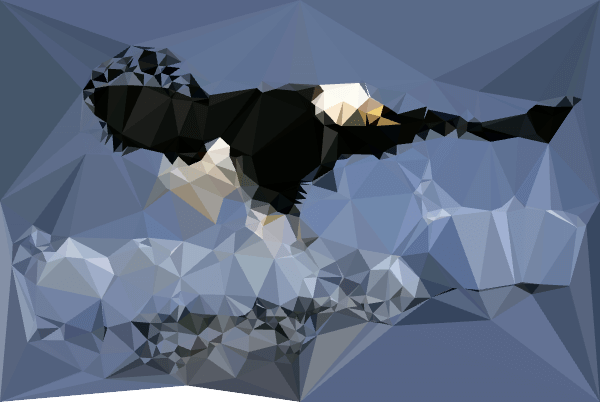}
 \caption{Two more examples created by our {\em TRIM} algorithm and Dmesh \cite{dmesh}. Our coarse triangulations capture the important features and closely resemble the original images. Left: The input images. Middle: The triangulations by {\em TRIM}. Right: The triangulations by DMesh \cite{dmesh}.}
 \label{fig:triangulation_more}
\end{figure}

\begin{figure}[t]
 \centering
 \includegraphics[width=0.24\textwidth]{bear_1}
 \includegraphics[width=0.24\textwidth]{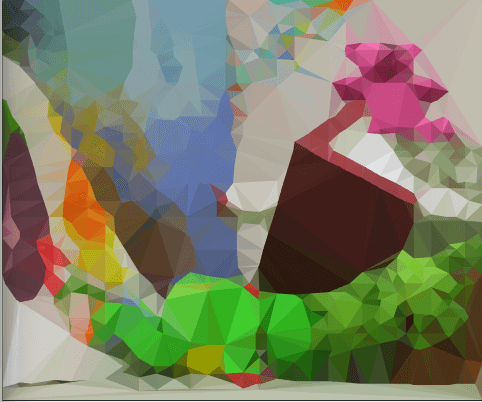}
 \includegraphics[width=0.24\textwidth]{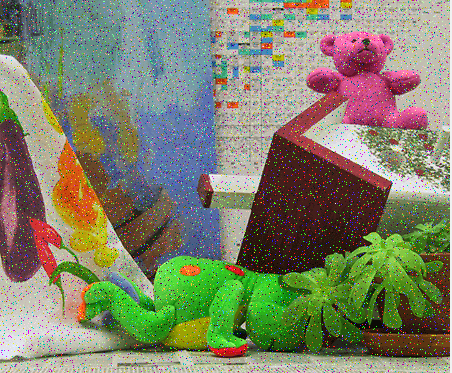}
 \includegraphics[width=0.24\textwidth]{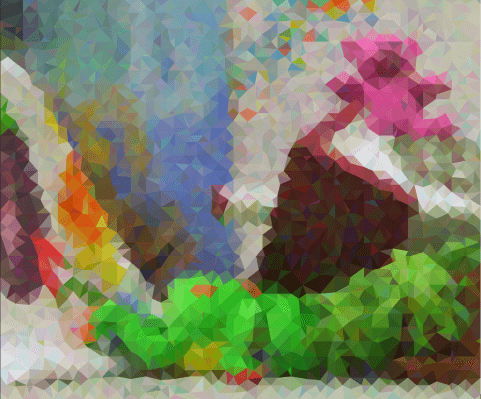}
 \includegraphics[width=0.24\textwidth]{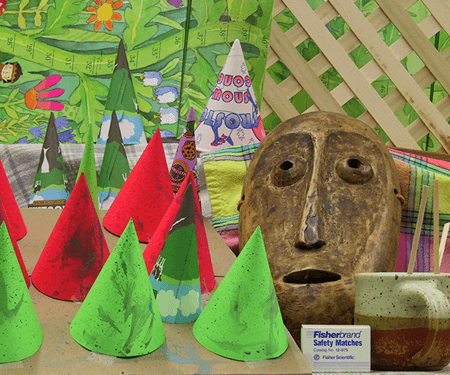}
 \includegraphics[width=0.24\textwidth]{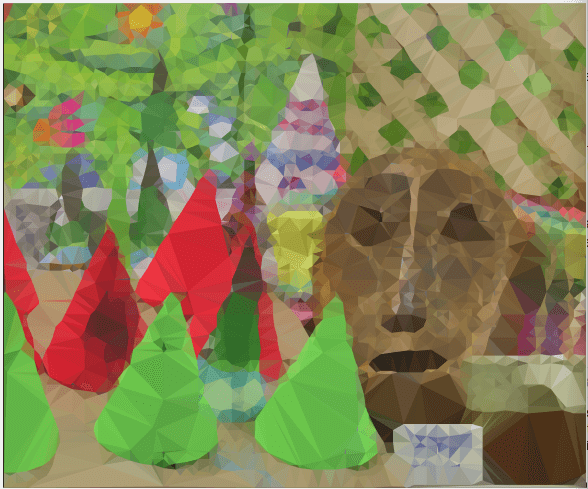}
 \includegraphics[width=0.24\textwidth]{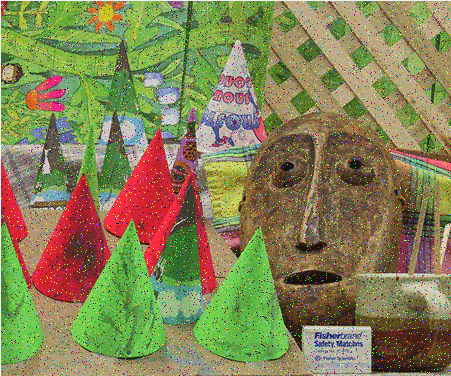}
 \includegraphics[width=0.24\textwidth]{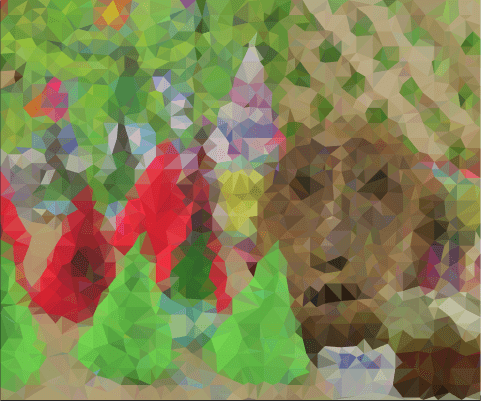}
 \caption{Two triangulation examples by our {\em TRIM} algorithm for noisy images. Left to right: The noise-free images, the triangulations computed by {\em TRIM} based on the noise-free images, the noisy images, and The triangulations computed by {\em TRIM} based on the noisy images. Note that the important features of the images are preserved even for noisy images.}
 \label{fig:noisy}
\end{figure}

\subsection{Performance of our proposed  triangulation (Algorithm \ref{algorithm:triangulation})}

In this subsection, we demonstrate the effectiveness of our triangulation scheme by various examples.

Our proposed algorithm is highly content-aware. Specifically, regions with high similarities or changes in color on an image can be easily recognized. As a result, the triangulations created faithfully preserve the important features by a combination of coarse triangles with different sizes. Some of our triangulation results are illustrated in Figure \ref{fig:triangulation}. For better visualizations, we color the resulting triangulations by mean of the original colors of corresponding patches. In Figure \ref{fig:triangulation_bee}, we apply our {\em TRIM} algorithm on a bee image. It can be observed that the regions of the green background can be effectively represented by coarser triangulations, while the region of the bee and flowers with apparent color differences is well detected and represented by a denser triangulation. Figure \ref{fig:triangulation_butterfly} shows another example of our triangulation result. The butterfly and the flowers are well represented in our triangulation result. The above examples demonstrate the effectiveness of our triangulation scheme for representing images in a simplified but accurate way. Some more triangulation examples created by our {\em TRIM} algorithm are shown in Figure \ref{fig:triangulation_more}. Figure \ref{fig:noisy} shows some triangulation examples for noisy images. It can be observed that our {\em TRIM} algorithm can effectively compute content-aware coarse triangulations even for noisy images.

We have compared our algorithm with the DMesh triangulator \cite{dmesh} in Figure \ref{fig:triangulation_bee}, Figure \ref{fig:triangulation_butterfly} and Figure \ref{fig:triangulation_more}. It can be observed that our triangulation scheme outperforms DMesh \cite{dmesh} in terms of the triangulation quality. Our results can better capture the important features of the images. Also, the results by DMesh \cite{dmesh} may contain unwanted holes while our triangulation results are always perfect rectangles. The comparisons reflect the advantage of our coarse triangulation scheme.

To quantitatively compare the content-aware property of our {\em TRIM} method and the DMesh method, we calculate the average absolute intensity difference $\frac{1}{N} \left\|I_{\text{triangulated}} - I_{\text{original}}\right\|_1$ between the original image $I_{\text{original}}$ (e.g. the left images in Figure \ref{fig:triangulation_more}) and the triangulated image $I_{\text{triangulated}}$ with piecewise constant color for each method (e.g. the middle and the right images in Figure \ref{fig:triangulation_more}), where $N$ is the number of pixels of the image. Table \ref{table:content_aware} lists the statistics. It is noteworthy that the average absolute intensity difference by {\em TRIM} is smaller than that by DMesh by around 30\% on average. This indicates that our TRIM algorithm is more capable to produce content-aware triangulations. 

\begin{table*}[h!]
\centering
\footnotesize
\begin{tabular}{|c|c|C{35mm}|C{35mm}|}
\hline
Image & Size & Average intensity difference ({\em TRIM}) & Average intensity difference (DMesh) \\ \hline
Bee & 640 $\times$ 425 & 0.1455 & 0.2115 \\ \hline
Bird & 1224 $\times$ 1224 & 0.1842 & 0.2074 \\ \hline
Butterfly & 1024 $\times$ 700 & 0.1629 & 0.2647\\ \hline
Book & 601 $\times$ 809 & 0.1446 & 0.2130 \\ \hline
Baseball & 410 $\times$ 399 & 0.1913 & 0.3554\\ \hline
Teddy & 450 $\times$ 375 & 0.1505 & 0.2998  \\ \hline
Pencil & 615 $\times$ 410 & 0.2610 & 0.4443 \\ \hline
Eagle & 600 $\times$ 402 & 0.1618 & 0.1897\\ \hline
\end{tabular}
\caption{The content-aware property of our {\em TRIM} algorithm and the DMesh method. }
\label{table:content_aware}
\end{table*}

\begin{figure}[t!]
 \centering
 \includegraphics[width=0.55\textwidth]{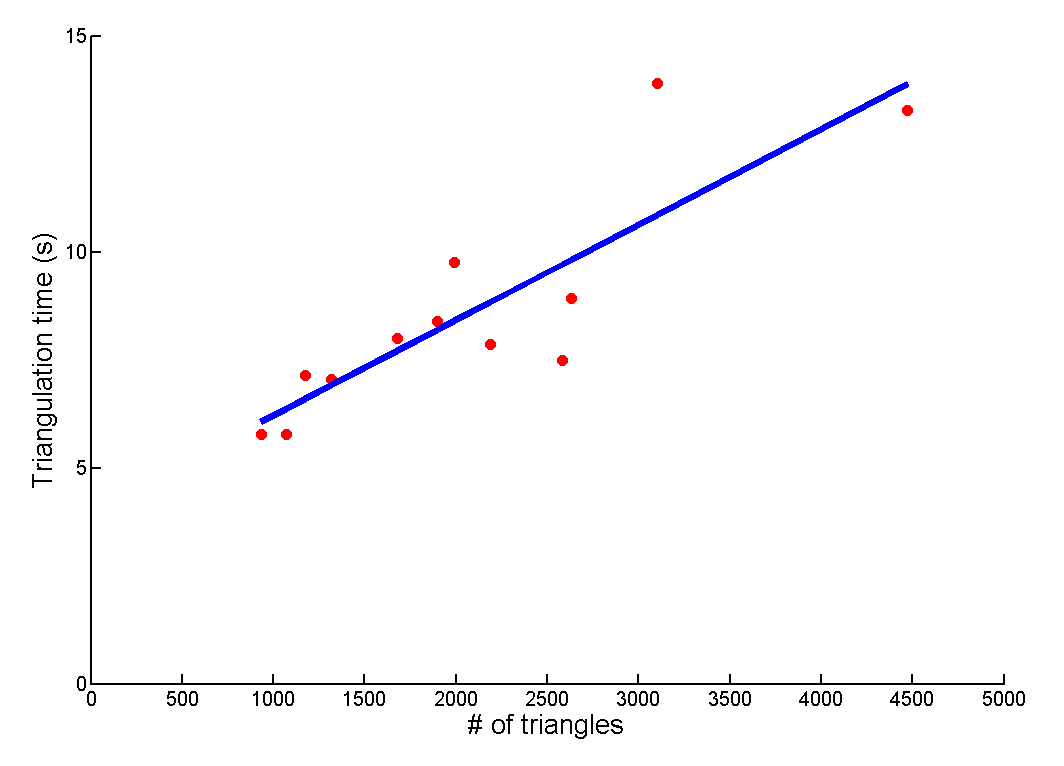}
 \caption{The relationship of the desired coarse triangulation size and the computational time of our proposed {\it TRIM} algorithm.}
 \label{fig:triangulation_time}
\end{figure}

Then, we evaluate the efficiency of our triangulation scheme for various images. Table \ref{table:trim} shows the detailed statistics. The relationship between the target coarse triangulation size and the computational time is illustrated in Figure \ref{fig:triangulation_time}. Even for high resolution images, the computational time for the triangulation is only around 10 seconds. It is noteworthy that our {\em TRIM} algorithm significantly compresses the high resolution images as coarse triangulations with only several thousand triangles.

\begin{table*}[h!]
\centering
\footnotesize
\begin{tabular}{|c|c|c|c|c|}
\hline
Image & Size & Triangulation time (s) & \# of triangles & Compression rate \\ \hline
Surfer & 846 $\times$ 421 & 5.78 & 1043 & 0.1536\% \\ \hline
Helicopter & 720 $\times$ 405 & 5.78 & 1129 & 0.1989\% \\ \hline
Bee & 640 $\times$ 425 & 7.13 & 1075 & 0.2029\% \\ \hline
Bird & 1224 $\times$ 1224 & 7.04 & 1287 & 0.0859\% \\ \hline
Butterfly & 1024 $\times$ 700 & 8.00 & 1720 & 0.1232\% \\ \hline
Book & 601 $\times$ 809 & 8.38 & 1629 & 0.3350\% \\ \hline
Baseball & 410 $\times$ 399 & 7.85 & 2315 & 0.7201\% \\ \hline
Teddy & 450 $\times$ 375 & 7.48 & 2873 & 0.8652\% \\ \hline
Pencil & 615 $\times$ 410 & 8.93 & 2633 & 0.5838\% \\ \hline
Tiger & 2560 $\times$ 1600 &  13.91 & 3105 & 0.0414\% \\ \hline 
Eagle & 600 $\times$ 402 & 13.27 & 1952 & 0.4299\% \\ \hline
\end{tabular}
\caption{Performance of our {\em TRIM} algorithm. The compression rate is $\frac{\text{\# of triangle nodes}}{\text{\# of pixels}} \times 100\%$.}
\label{table:trim}
\end{table*}

\begin{figure}[t!]
 \centering
 \includegraphics[width=0.24\textwidth]{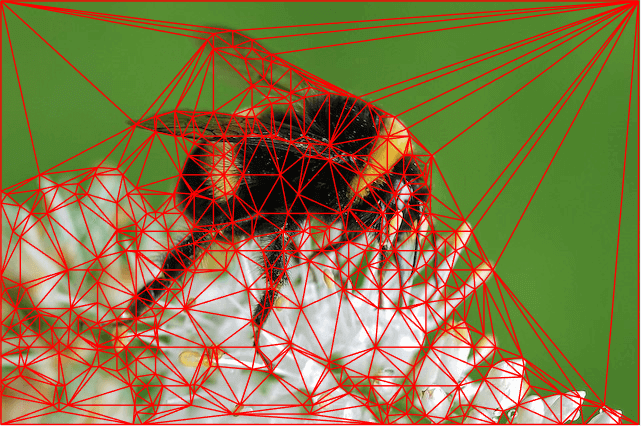}
 \includegraphics[width=0.24\textwidth]{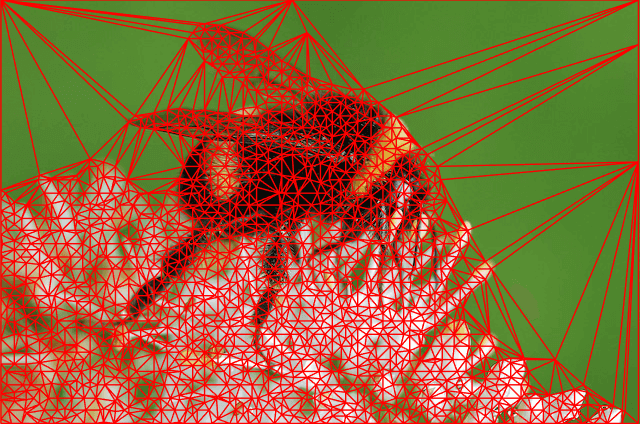}
 \includegraphics[width=0.235\textwidth]{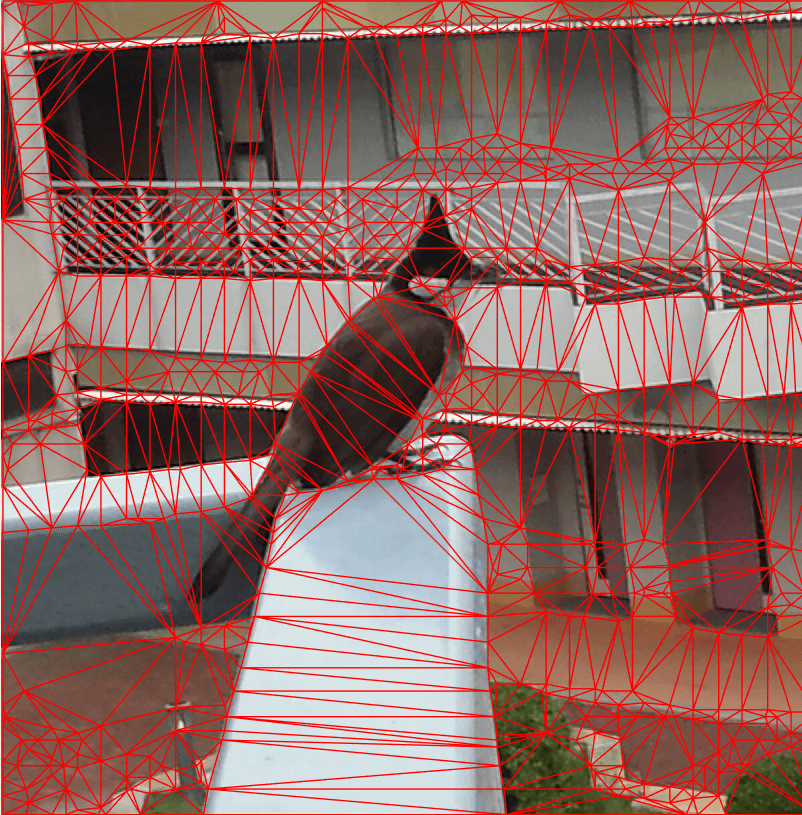}
 \includegraphics[width=0.24\textwidth]{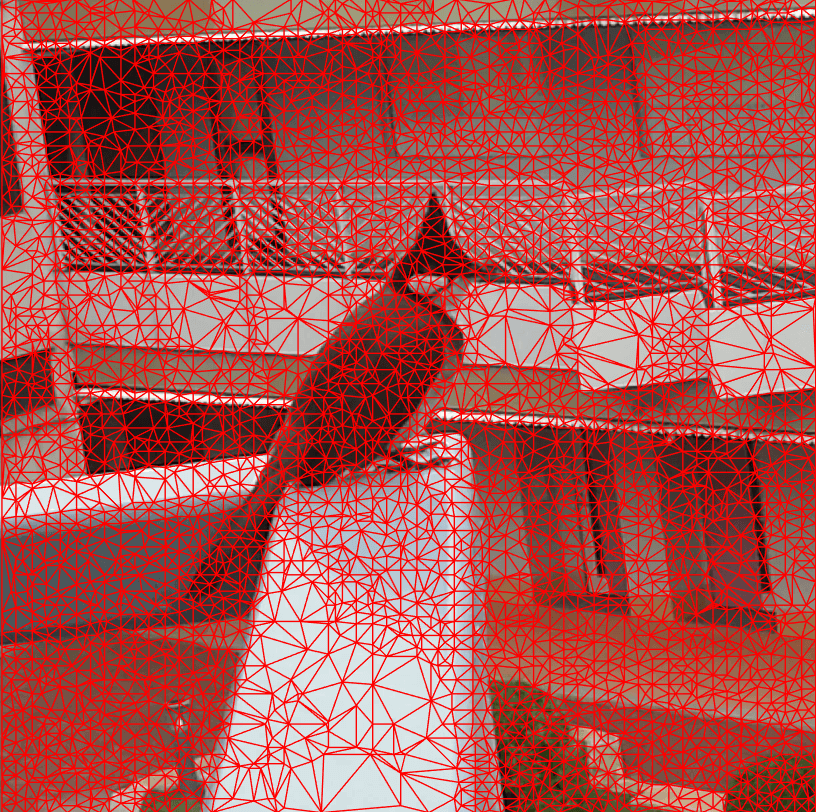}
 \caption{The triangulations created by our {\em TRIM} algorithm (left) and by the algorithm without the segmentation step (Right). The results show that the segmentation step is crucial for achieving a coarse triangulation. Number of triangles produced (left to right): 923, 3612, 1496, 8685.}
 \label{fig:segmentation}
\end{figure}

It is noteworthy that the combination of the steps in our {\em TRIM} algorithm is important for achieving a coarse triangulation. More specifically, if certain steps in our algorithm are removed, the triangulation result will become unsatisfactory. Figure \ref{fig:segmentation} shows two examples of triangulations created by our entire {\em TRIM} algorithm and by our algorithm with the segmentation step excluded. It can be easily observed that without the segmentation step, the resulting triangulations are extremely dense and hence undesirable for simplifying further computations. By contrast, the number of triangles produced by our entire {\em TRIM} algorithm is significantly reduced. The examples highlight the importance of our proposed combination of steps in the {\em TRIM} algorithm for content-aware coarse triangulation.

\begin{figure}[t!]
 \centering
 \includegraphics[width=0.9\textwidth]{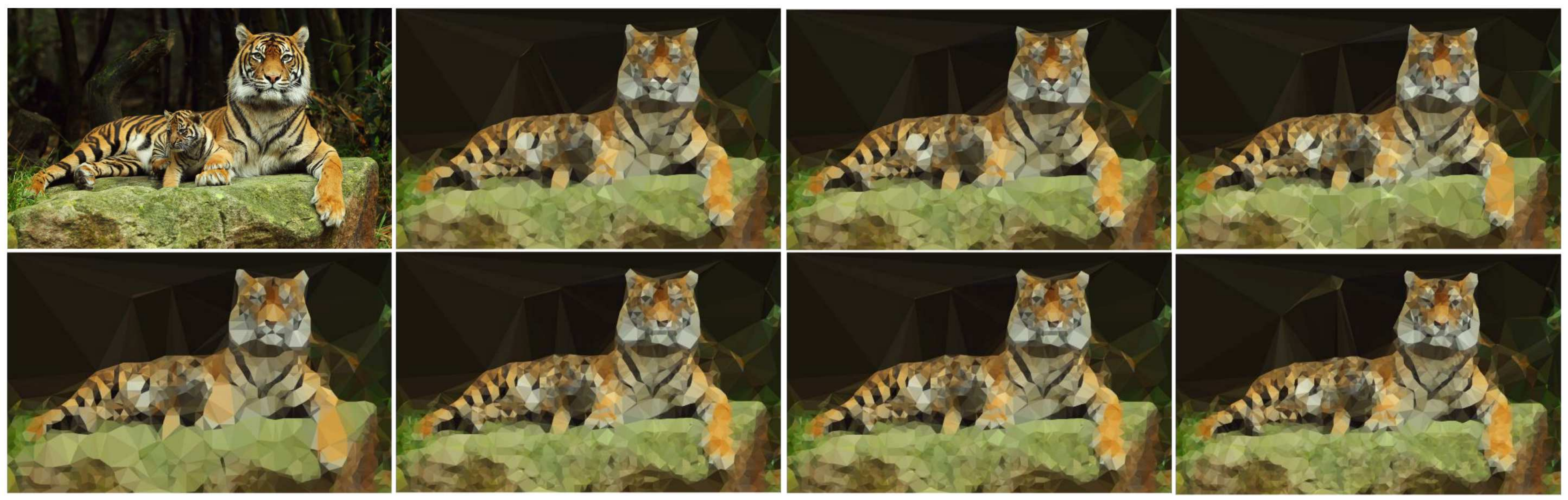}
 
\centering
\begin{tabular}{|C{17mm}|c|c|c|c|c|c|c|}
\hline
$\lambda$ & 0.5 & 0.25 & 0.75 & 0.4 & 0.7 & 0.2 & 1 \\ \hline
$\sigma$,$s$ & 2 & 1.75 & 2.25 & 2.5 & 3 & 1.5 & 2\\ \hline
$\theta$ & 0.5 & 0.25 & 0.4 & 0.6 & 0.3 & 0.2 & 0 \\ \hline
Average intensity difference & 0.2227 & 0.2220 & 0.2233 & 0.2306 & 0.2195 & 0.2166 & 0.2124 \\ \hline
\end{tabular}
 \caption{The triangulation results with different parameters $(\lambda,\sigma,s,\theta)$ in the unsharp masking step. Top left: The original image. Top middle left to bottom right: results with different parameters.)}
 \label{fig:unsharp_parameters}
\end{figure}

We also analyze the sensitivity of the triangulated images to the parameters in the unsharp masking step. Figure \ref{fig:unsharp_parameters} shows several triangulation results with different choice of $(\lambda,\sigma,s,\theta)$. It can be observed that the triangulation results are robust to the parameters.

\subsection{Registration of two triangulated image surfaces (Algorithm \ref{algorithm:registration})}

\begin{figure}[t!]
 \centering
 \includegraphics[width=0.7\textwidth]{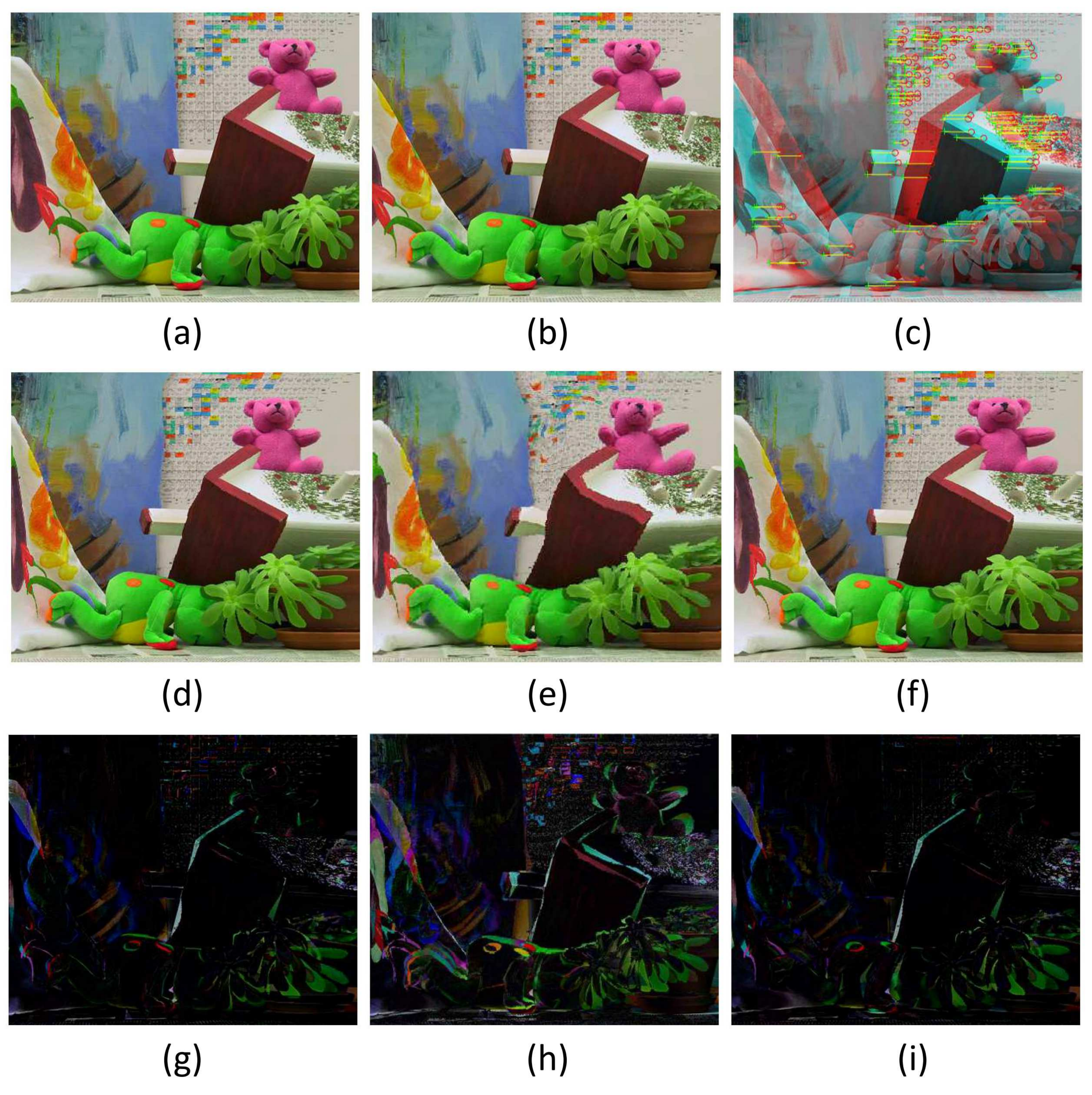}
 \caption{Stereo landmark registration of two images of doll using different algorithms. (a): The source image. (b): The target image. (c): The prescribed feature correspondences. (d): The registration result by the dense grid-based approach (4 pixels per grid). (e): The registration result via DMesh \cite{dmesh}. (f): The registration result by our {\em TRIM}-based method. (g): The intensity difference after the registration by the dense grid-based approach. (h): The intensity difference after the registration via DMesh \cite{dmesh}. (i): The intensity difference after the registration by our {\em TRIM}-based method.}
 \label{fig:registration_teddy}
\end{figure}

\begin{figure}[t!]
 \centering
 \includegraphics[width=0.7\textwidth]{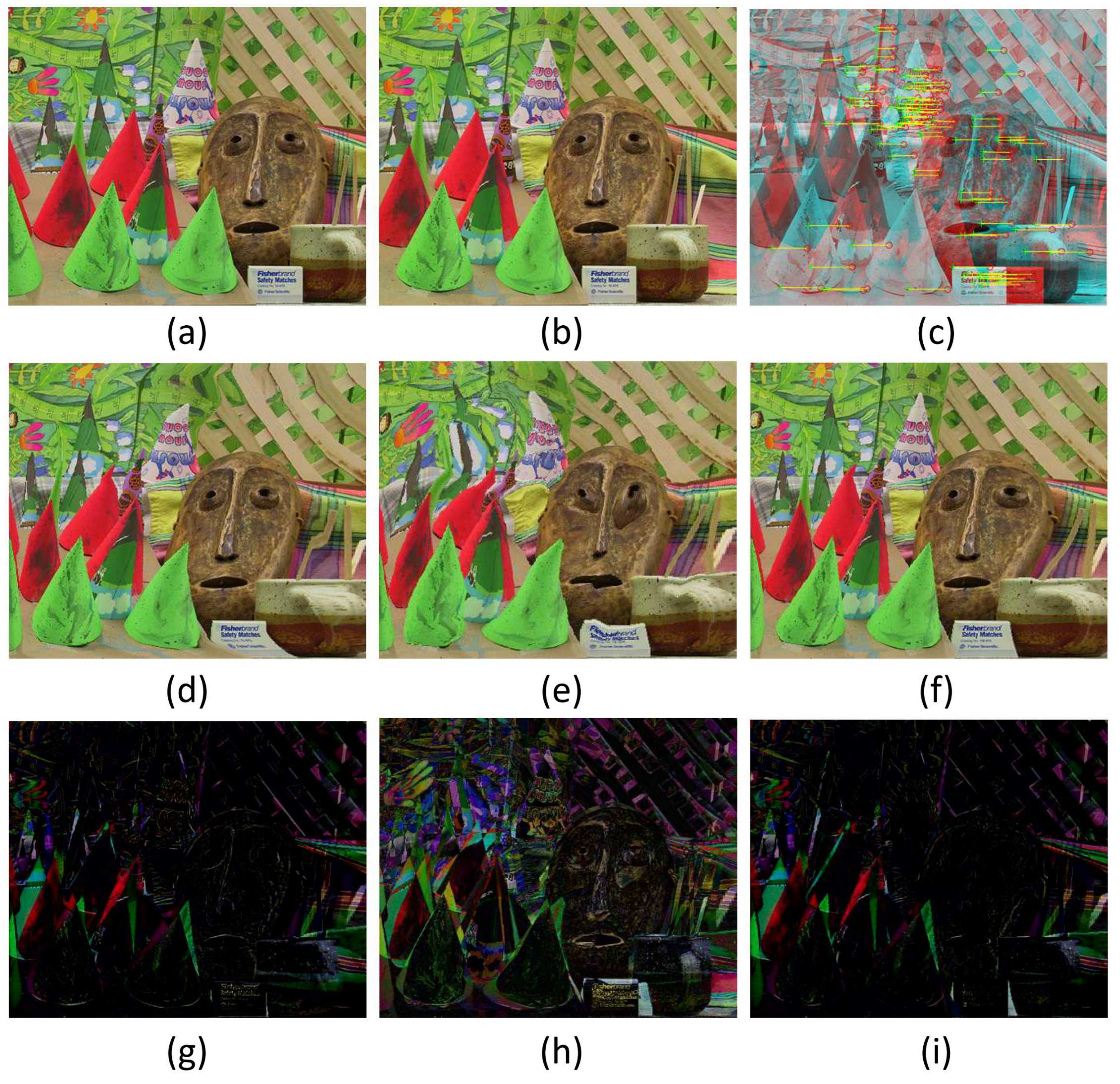}
 \caption{Stereo landmark registration of two images of cones using different algorithms. Please refer to Figure \ref{fig:registration_teddy} for the description of the images.} 
 \label{fig:registration_cone}
\end{figure}

In this subsection, we demonstrate the effectiveness of our proposed triangulation-based method for landmark-based image registration. In our experiments, the feature points on the images are extracted using the Harris--Stephens algorithm \cite{harris88} as landmark constraints. The landmark extraction is fully automatic. More specifically, we use the MATLAB functions \texttt{detectHarrisFeatures},\texttt{extractFeatures} and \texttt{matchFeatures} on the images. For the teddy example, 132 landmark pairs are generated using the above procedure. For the cones example, 162 landmark pairs are generated.

For simplifying the image registration problems, one conventional approach is to make use of coarse regular grids followed by interpolation. It is natural to ask whether our proposed coarse triangulation-based method produces better results. In Figure \ref{fig:registration_teddy}, we consider a stereo registration problem of two scenes. With the prescribed feature correspondences, we compute the feature-endowed stereo registration via the conventional grid-based approach, the DMesh triangulation approach \cite{dmesh} and our proposed {\it TRIM} method. For the grid-based approach and the DMesh triangulation approach \cite{dmesh}, we take the mesh vertices nearest to the prescribed feature points on the source image as source landmarks. For our proposed {\it TRIM} method, as the landmark vertices are automatically embedded in the content-aware coarse triangulation, the source landmarks are exactly the feature points detected by the method in \cite{harris88}.

It can be observed that our triangulation-based approach produces a much more natural and accurate registration result when compared with both the grid-based approach and the DMesh triangulation approach. In particular, sharp features such as edges are well preserved using our proposed method. By contrast, the edges are seriously distorted in the other two methods. In addition, the geometry of the background in the scenes are well retained via our {\em TRIM} method but not the other two methods. The higher accuracy of the registration result by our approach can also be visualized by the intensity difference plots. Our triangulation-based approach results in an intensity difference plot with more dark regions than the other two approaches. The advantage of our method over the other two methods is attributed to the geometry preserving feature of our {\it TRIM} algorithm, in the sense that the triangulations created by {\it TRIM} are more able to fit into complex features and have more flexibilities in size than regular grids. Also, the triangulations created by DMesh \cite{dmesh} do not capture the geometric features and hence the registration results are unsatisfactory. They reflect the significance of our content-aware {\it TRIM} triangulation scheme in computing image registration. Another example is illustrated in Figure \ref{fig:registration_cone}.
Again, it can be easily observed that our proposed {\em TRIM} triangulation approach leads to a more accurate registration result.

To highlight the improvement in the efficiency by our proposed {\em TRIM} algorithm, Table \ref{table:registration} records the computational time and the error of the registration via the conventional grid-based approach and our {\it TRIM} triangulation-based approach. It is noteworthy that our proposed coarse triangulation-based method significantly reduces the computational time by over 85\% on average when compared with the traditional regular grid-based approach. To quantitatively assess the quality of the registration results, we define the matching accuracy by
\begin{equation}
A = \frac{\text{\# pixels for which} \|\text{final intensity - original intensity}\|_1 \text{ is less than } \epsilon}{\text{Total \# of pixels}} \times 100\%.
\end{equation}
The threshold $\epsilon$ is set to be $0.2$ in our experiments. Our triangulation-based method produces registration results with the matching accuracy higher than that of the regular grid-based method by 6\% on average. The experimental results reflect the advantages of our {\it TRIM} content-aware coarse triangulations for image registration.

\begin{table*}[h!]
\centering
\footnotesize
\begin{tabular}{|c|c|C{12mm}|C{19mm}|C{12mm}|C{19mm}|C{18mm}|}
\hline
Images & Size & \multicolumn{4}{c|}{Registration} & Time saving rate \\ \cline{3-6}
& & \multicolumn{2}{c|}{Via regular grids} & \multicolumn{2}{c|}{Via {\em TRIM}} & \\ \cline{3-6}
& & Time (s) & Matching accuracy (\%) & Time (s) & Matching accuracy (\%) & \\ \hline
Teddy & 450 $\times$ 375 & 102.3 & 59.5 & 13.8 & 70.7 & 86.5103\% \\ \hline
Cones & 450 $\times$ 375 & 108.7 & 51.3 & 28.2 & 61.2 & 74.0570\%\\ \hline
Cloth & 1252 $\times$ 1110 & 931.0 & 70.7 & 36.0 & 75.4 & 96.1332\% \\ \hline
Books & 1390 $\times$ 1110 & 1204.5 & 59.0 & 51.0 & 63.0 & 95.7659\% \\ \hline
Dolls & 1390 $\times$ 1110 & 94.3 & 62.3 & 11.0 & 62.3 & 88.3351\% \\ \hline
\end{tabular}
\caption{The performance of feature-based image registration via our proposed {\em TRIM} coarse triangulation method and the ordinary coarse grids. Here, the time saving rate is defined by $\frac{\text{Registration time via regular grids} - \text{Registration time via {\em TRIM}}}{\text{Registration time via regular grids}} \times 100\%$.}
\label{table:registration}
\end{table*}

We further compare our {\it TRIM}-based registration method with two other state-of-the-art image registration methods, namely the Large Displacement Optical Flow (LDOF) \cite{Brox11} and the Diffeomorphic Log-Demons \cite{Lombaert14}. Table \ref{table:registration_comparison} lists the performance of the methods. It is noteworthy that our method is significantly faster than the two other methods, with at least comparable and sometimes better matching accuracy.

\begin{table*}[h!]
\centering
\footnotesize
\begin{tabular}{|c|c|C{10mm}|C{15mm}|C{10mm}|C{15mm}|C{10mm}|C{15mm}|}
\hline
Images & Size & \multicolumn{2}{c|}{{\em TRIM}} & \multicolumn{2}{c|}{LDOF} & \multicolumn{2}{c|}{Spectral Log-Demons} \\ \cline{3-8}
& & Time (s) & Matching accuracy (\%) & Time (s) & Matching accuracy (\%) & Time (s) & Matching accuracy (\%)  \\ \hline
Aloe & 222 $\times$ 257 & 2.8 & 91.1 & 12.1 & 86.4 & 18.2 & 94.6  \\ \hline
Computer & 444 $\times$ 532 & 4.0 & 69.9 & 51.4 & 70.0 & 7.9 & 41.3  \\ \hline
Laundry & 444 $\times$ 537 & 4.1 & 73.2 & 52.4 & 75.7 & 9.02 & 50.8  \\ \hline
Dwarves & 777 $\times$ 973 & 7.9 & 80.5 & 311.8 & 82.5 & 36.7 & 50.8  \\ \hline
Art & 1390 $\times$ 1110 & 13.5 & 84.8 & 1110.6 & 87.4 & 242.9 & 77.9  \\ \hline
Bowling2 & 1110 $\times$ 1330 & 20.9 & 90.1 & 1581.9 & 86.1 & 22.0 & 57.6  \\ \hline
\end{tabular}
\caption{Comparison between our {\em TRIM}-based image registration, the Large Displacement Optical Flow (LDOF) \cite{Brox11} and the Spectral Log-Demons \cite{Lombaert14}.}
\label{table:registration_comparison}
\end{table*}

\begin{figure}[t!]
 \centering
 \includegraphics[width=0.9\textwidth]{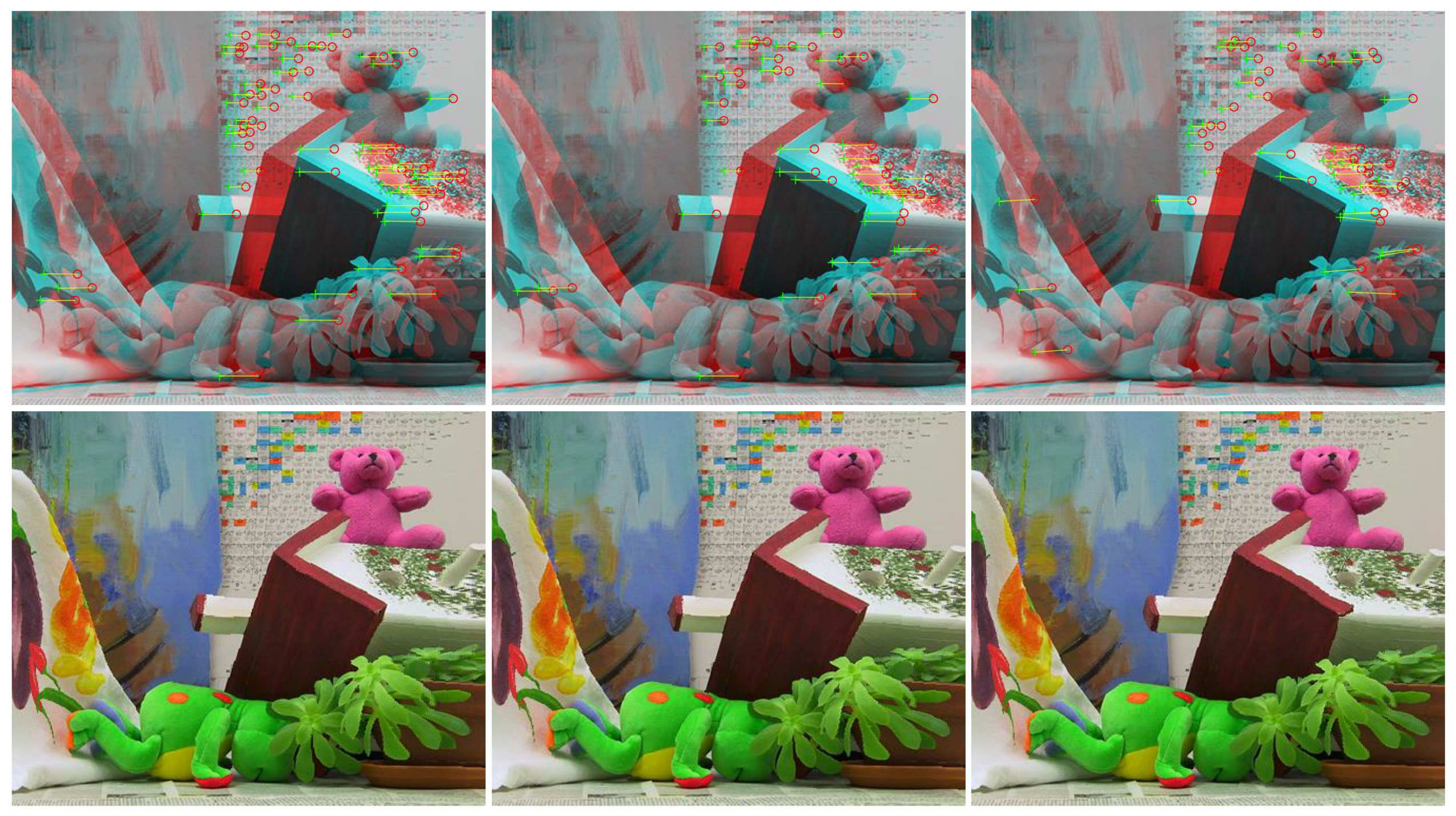}
 \caption{Different landmark correspondences and the \emph{TRIM}-based registration results for the teddy example in Figure \ref{fig:registration_teddy}. Left: Only 1/2 of the landmark pairs in Figure \ref{fig:registration_teddy} are randomly selected for computing the registration. Middle: Only 1/3 of the landmark pairs are used. Right: Only 1/3 of the landmark pairs are used, with 2\% random noise added to the target landmark locations. The matching accuracies are respectively $70.5\%, 70.4\%, 67.2\%$, which are very close to the original result ($70.7\%$).} 
 \label{fig:landmark_stability}
\end{figure}

Besides, we study the stability of the \emph{TRIM}-based registration result with respect to the feature points detected. Figure \ref{fig:landmark_stability} shows the results with different feature correspondences, including a change in the number of landmark pairs and a change in the target landmark position. From the resulting triangulated images and the statistics on the matching accuracy, it can be observed that the deformation is stable with respect to the choice of the feature points.

\section{Conclusion and future work} \label{conclusion}
In this paper, we have proposed a new image registration algorithm (Algorithm \ref{algorithm:registration}), which operates on content-aware coarse triangulations to aid registration of high resolution images. The obtained algorithm is computationally efficient and capable to achieve a highly accurate result while resembling the original image. It has two stages with stage $1$ obtaining content-aware coarse triangulations and stage $2$ registering two triangulated surfaces. Both algorithms can be used as standalone methods: Algorithm \ref{algorithm:triangulation} for extracting main features of images (compression) and Algorithm \ref{algorithm:registration} for registering two surfaces (quality mapping).

 Our proposed method is advantageous for a large variety of registration applications with a significant improvement of the computational efficiency and registration accuracy. Our proposed method can also serve as an effective initialization for other registration algorithms. In the future, we aim to extend our proposed algorithm to high dimensions.

\end{document}